\documentclass[prd,notitlepage,nofootinbib,preprintnumbers,amssymb,superscriptaddress]{revtex4-1}
\usepackage{amsfonts,amssymb,amsmath,mathrsfs,graphicx,color,bm,ulem,mathtools}
\usepackage{latexsym,subfigure,ascmac}
\usepackage{here}
\usepackage{braket}
\definecolor{ultramarine}{rgb}{0.07, 0.04, 0.56}
\definecolor{cadmiumgreen}{rgb}{0.0, 0.42, 0.24}
\definecolor{indigo(dye)}{rgb}{0.0, 0.25, 0.42}
\usepackage[linktocpage=true]{hyperref}
\hypersetup{
colorlinks=true,
citecolor=ultramarine,
linkcolor=cadmiumgreen,
urlcolor=indigo(dye),
}

\newcommand{\f}[2]{\frac{#1}{#2}}  
\newcommand{\mk}[1]{\left( #1 \right)}  
\newcommand{\kk}[1]{\left[ #1 \right]}  
  
\newcommand{\be}{\begin{equation}}  
\newcommand{\ee}{\end{equation}}
\newcommand{\D}{\Delta}
\newcommand{\sep}{\lambda}

\newcommand{\df}{\dfrac}

\graphicspath{{Images/}}

\def\@fnsymbol#1{\ensuremath{\ifcase#1
\or $\textleaf$ \or $\PHplaneTree$ \or $\PHrosette$ \or $\PHvine$
\else\@ctrerr\fi}}%

\makeatother

\def\sideremark#1{\ifvmode\leavevmode\fi\vadjust{\vbox to0pt{\vss
 \hbox to 0pt{\hskip\hsize\hskip1em
 \vbox{\hsize1.5cm\tiny\raggedright\pretolerance10000
 \noindent #1\hfill}\hss}\vbox to8pt{\vfil}\vss}}}%
\def\@fnsymbol#1{\ensuremath{\ifcase#1
\or $\textleaf$ \or $\PHplaneTree$ \or $\PHrosette$ \or $\PHvine$
\else\@ctrerr\fi}}%

\begin{document}
\title{Spectroscopy of Kerr-AdS$_5$ spacetime with the Heun function:\\ Quasinormal modes, greybody factor, and evaporation}

\author{Sousuke \surname{Noda}}
\email{snoda@cc.miyakonojo-nct.ac.jp}
\affiliation{National Institute of Technology, Miyakonojo College, Miyakonojo 885-8567, Japan}

\author{Hayato \surname{Motohashi}}
\email{motohashi@cc.kogakuin.ac.jp}
\affiliation{Division of Liberal Arts, Kogakuin University, 2665-1 Nakano-machi, Hachioji, Tokyo, 192-0015, Japan}

\begin{abstract}
We investigate quasinormal modes, greybody factor, and Hawking evaporation of 
a five-dimensional Kerr-anti de Sitter (AdS$_5$) black hole by solving a wave equation for a test massive scalar field in terms of the local Heun function. 
We clarify the distribution of the quasinormal modes satisfying the ingoing and decaying boundary conditions at the event horizon and conformal infinity, respectively, and their splitting behaviors in the complex frequency plane with respect to a variation of spin parameters.
We also point out the existence of purely imaginary modes.
Further, we develop a method to extract in/outgoing waves near the AdS boundary, which enables us to define the greybody factor for a wide range of parameter space complementary to previous works.
Using the greybody factor, we compute evaporation rates for mass and spins for the Kerr-AdS$_5$ black hole via the Hawking radiation of the scalar field.
\end{abstract}

\maketitle  


\section{Introduction}
\label{sec:intro}

A black hole is a spacetime region surrounded by an event horizon. Although the event horizon itself is 
not observable, several black hole candidates 
have been observed starting by X-ray observation 
\cite{1972Natur.235...37W,1972Natur.235..271B}, gravitational wave detection \cite{Abbott:2016blz}, and imaging the horizon scale region around supermassive black holes \cite{EventHorizonTelescope:2019dse, EventHorizonTelescope:2022xnr}. 
In these observations, what to be detected were waves emitted and scattered around a black hole.
Theoretically, considering quantum effects, black holes themselves can also emit a Hawking radiation \cite{Hawking:1974rv,Hawking:1975vcx}.
These properties of a black hole can be investigated by the linear perturbation theory around the black hole spacetime. 
The separability of the variables of the master equation has been clarified in the Kerr background
\cite{Teukolsky:1972my,Teukolsky:1973ha,Unruh:1973bda,Chandrasekhar:1976ap}, 
in the Kerr-de Sitter background \cite{Khanal:1983vb, Chambers:1994ap}, and in the Kerr-anti de Sitter (AdS) background 
\cite{Chambers:1994ap, Giammatteo:2005vu, Cardoso:2006wa}. 
Besides four dimensional black holes, 
higher dimensional black holes are also interesting 
from the theoretical point of view since they have more
fruitful properties than the four dimensional ones. 
Furthermore, the study of higher dimensional Kerr-AdS black holes is important 
in the context of gauge/gravity correspondence 
\cite{Maldacena1997, Gubser1998, Witten1998}.
In general, by solving a wave equation in a black hole spacetime, it is possible to understand the properties of the 
black hole by quasinormal (QN) modes, wave 
scattering phenomena, evaporation rate via Hawking radiation. 
Such an approach is known as black hole spectroscopy~\cite{Maldacena1997}.

The QN modes of a black hole are damped oscillations of 
fields characterized by discrete complex frequencies given as poles of the Green function satisfying a certain set of boundary conditions.
Investigating the QN modes, one can discuss the 
late time behavior of a gravitational wave signal, stability, and quantum aspects 
of the black hole~\cite{Kokkotas:1999bd,Berti:2009kk,Konoplya:2011qq}.
The QN modes have also been studied for asymptotically AdS black holes in the context of the gauge/gravity correspondence to examine a holographic dual system \cite{Horowitz:1999jd,Nunez:2003eq,Kovtun:2005ev,Birmingham:2001pj}.
Therefore, study of the QN modes is important from both theoretical and observational point of view.

In general, several sequences of the QN modes show up for a black hole spacetime.
A peculiar sequence would be mode(s) that align on the negative imaginary axis, corresponding to pure diffusion.
The most well-known example is so-called algebraically special mode of the Schwarzschild spacetime \cite{Couch:1973zc, Chandrasekhar:1984}.
The Kerr spacetime also admits the purely imaginary modes \cite{Yang:2013uba,Hod:2013fea}.
There is some subtlety on these modes, and in general they are not always the QN modes~\cite{MaassenvandenBrink:2000iwh,Cook:2016fge,Cook:2016ngj} (see also Appendix~A in \cite{Berti:2009kk}). 
The purely imaginary modes are also studied for 
the Schwarzschild spacetime~\cite{Khriplovich:2005wf,Cho:2005yc,Mamani:2022akq},
the Schwarzschild-AdS$_4$ spacetime \cite{Cardoso:2001bb,Cardoso:2003cj}, Reissner-Nordstr\"{o}m-AdS$_4$ spacetime \cite{Katagiri:2020mvm}, 
Ba\~nados-Teitelboim-Zanelli spacetime \cite{Katagiri:2022qje}, and for D3/D7 brane system \cite{Ishigaki:2021vyv,Atashi:2022ufl}.
They are also known as the non-hydrodynamic modes~\cite{Amado:2009ts,Finazzo:2016psx,Mamani:2018qzl,Mamani:2022qnf}.
They are often discussed in the context of gauge/gravity correspondence as well and are expected to play an important role \cite{Cardoso:2001bb,Kovtun:2005ev,Ishigaki:2021vyv}.
The QN modes and the superradiant instability of Kerr-AdS$_5$ spacetime have been extensively investigated \cite{Murata2009, Cardoso:2013pza, BarraganAmado:2018zpa, Amado:2021erf, Koga:2022vun}, 
and it was clarified that there exist QN modes whose real parts approach zero in the ultra-spinning limit \cite{Koga:2022vun}.
However, to the best of our knowledge, it has not been clarified yet whether the purely imaginary modes exist for the Kerr-AdS$_5$ spacetime including lower spin case.

On the other hand, besides the QN modes, Hawking radiation is also an important feature of black holes. 
Originally, it was discussed for asymptotically flat black holes, but later it was generalized to asymptotically AdS case \cite{Hawking:1982dh}. 
Although black holes emit Hawking radiation, some of them are scattered back to the black hole horizon due to the potential barrier formed by 
the gravitational force and angular momenta of the emitted radiations. Hence, for black holes to evaporate via Hawking radiation, emitted radiations
need to pass through the potential barrier.
The transmittance is called greybody factor \cite{Page:1976df,Page:1976ki,Page:1977um}, which  
corresponds to
the efficiency of the evaporation of the black hole.
The definition and computation of the greybody factor for asymptotically flat or de Sitter black holes is straightforward since the asymptotic structure of these spacetimes admits the purely outgoing boundary condition for the fields.
However, for the asymptotically AdS spacetimes, 
the situation is more subtle
due to the existence of the AdS boundary which plays a role of a box to confine those fields. 
As a specific problem in defining wave scattering, one should be careful to identify ingoing and outgoing waves around the outer potential barrier.
Regarding this difficulty, a direct computation of the greybody factor for the asymptotically AdS black holes has been addressed in \cite{Harmark:2007jy,Rocha:2009xy,Jorge:2014kra}, and the absorption cross section via gauge/gravity correspondence has also been discussed in \cite{Teo:1998dw,Muller-Kirsten:1998ijv}. 
The computation of the greybody factor considered in the previous 
works is based on the approximated solution of the  
equation of motion, and hence applies to a limited parameter region. 
On the other hand, the evaluation with the exact solution in terms of the Heun function may enable us 
to obtain the greybody factor for a wider range of parameters, and to compute the evaporation rates properly.

The Heun's equation is a second-order linear differential equation possessing four regular singular points \cite{Heun1889,ronveaux1995heun,slavianov2000special,Maier_2006}.
It is known that the Teukolsky equation for the four-dimensional Kerr-Newman-de Sitter spacetime can be transformed into the Heun's equation \cite{Suzuki:1998vy}. 
The exact solution is known as local Heun function or general Heun function.
Recently, it has been employed to the computation of QN modes~\cite{Hatsuda:2020sbn,Oshita:2021iyn}, wave scattering problem and the Green's function~\cite{Motohashi:2021zyv}, the greybody factor~\cite{CarneirodaCunha:2015qln,Novaes:2018fry}, 
and Hawking radiation~\cite{Gregory:2021ozs,Nambu:2021eqe,Oshita:2022pyf}.
Since the Klein-Gordon equation for a test scalar field for an asymptotically AdS black holes can also be transformed into the Heun's equation if the spacetime dimension is five~\cite{Aliev:2008yk,Amado:2017kao,BarraganAmado:2018zpa}, it is interesting to investigate the Kerr-AdS$_5$ with the exact solution in terms of the local Heun function.

In this paper, we investigate the QN modes, greybody factor, and Hawking evaporation of a Kerr-AdS$_5$ black hole for a 
massive scalar field by using the exact solution of the Klein-Gordon equation in terms of the local Heun function. 
In addition to known sequences of the QN modes, we find that the purely imaginary modes exist for the Kerr-AdS$_5$ spacetime.
For the computation of the greybody factor, we develop a method to extract in/outgoing waves near the AdS boundary, which allows us to explore the Hawking radiation for a parameter region complementary to the previous work.

The rest of the present paper is organized as follows. 
In \S\ref{sec:sc}, we introduce a test massive scalar field in the Kerr-AdS$_5$ spacetime, and demonstrate the transformation of both the radial and angular equations to the Heun equation. While the previous work focused on the case where magnetic quantum numbers are non-negative, we consider all the possible cases.
Then, we derive formulae for coefficients connecting local Heun functions at each regular singular point. As an application of those formulae, we make a comment on the Gubser-Klebanov-Polyakov(GKP)-Witten relation \cite{Gubser1998,Witten1998}, which is a 
dictionary in the gauge/gravity correspondence \cite{Maldacena1997}. 
In \S\ref{sec:qnm}, we investigate the QN modes, and 
show that nonzero spin parameter breaks the degeneracy between the QN modes, causing a Zeeman-like splitting behavior depending on the value of magnetic quantum numbers.
Further, we point out the existence of the purely imaginary modes with vanishing total magnetic quantum number.
In \S\ref{sec:greybody}, we develop a method to distinguish in/outgoing mode near the AdS boundary by using the local Heun function, and provide the greybody factor.
We then calculate the evaporation rates of the mass and angular momenta of the Kerr-AdS$_5$ black hole in \S\ref{sec:evap}, focusing on the contribution from the Hawking radition of the scalar field.
\S\ref{sec:con} is devoted to conclusion. 
Throughout the present paper, we use the natural units: $c=G=\hbar=k_\text{B}=1$.

\section{Klein-Gordon equation in 
Kerr-AdS$_5$ spacetime}
\label{sec:sc}
The metric of the Kerr-AdS$_5$ spacetime is given as \cite{Myers:1986un,Hawking:1998kw,Gibbons:2004ai}
\begin{align}
    ds^2&=-\df{\D}{\rho^2}\mk{dt-\df{a_1 \sin^2\theta}{\Xi_1}d\phi_1-\dfrac{a_2 \cos^2\theta}{\Xi_2}}^2+\df{\rho^2}{\D}dr^2+\df{\rho^2}{\D_\theta}d\theta^2 + \df{\D_\theta \sin^2\theta}{\rho^2}\mk{a_1 dt -\df{r^2+a_1^2}{\Xi_1} d\phi_1}^2\\
    &+\df{\D_\theta \cos^2\theta}{\rho^2}\mk{a_2 dt -\df{r^2+a_2^2}{\Xi_2} d\phi_2}^2+\df{1+r^2/L^2}{r^2\rho^2}\mk{a_1 a_2 dt -\df{a_2(r^2+a_1^2)\sin^2\theta}{\Xi_1}d\phi_1 -\df{a_1(r^2+a_2^2)\cos^2\theta}{\Xi_2}d\phi_2}^2,
\end{align}
where $L$ is the AdS radius and
\begin{align}
    &\D\coloneqq \frac{1}{r^2}(r^2+a_1^2)(r^2+a_2^2)\mk{1+\df{r^2}{L^2}}-2M=\df{1}{r^2L^2}(r^2-r_0^2)(r^2-r_+^2)(r^2-r_-^2),\\
   &\D_\theta\coloneqq 1-\df{a_1^2}{L^2}\cos^2\theta-\df{a_2^2}{L^2}\sin^2\theta,\quad
    \rho^2\coloneqq r^2+a_1^2\cos^2\theta +a_2^2\sin^2\theta,\\
   &\Xi_1\coloneqq 1-\df{a_1^2}{L^2},\quad \Xi_2\coloneqq 1-\df{a_2^2}{L^2}.
\end{align}
Here, $M$ is a mass parameter, $a_1,a_2$ are spin parameters, and 
$r_k\ (k=+,-,0)$ represent horizon radii.
We assume that they are distinct and $r_0^2<0<r_-^2<r_+^2$.
Namely, $r_0$ is pure imaginary, and $r_-,r_+$ are real and satisfy $0<r_-<r_+$.
Then $r_-,r_+$ are radii of the inner and outer horizons, respectively.
It holds that 
\begin{align} 
\label{crel1} &r_0^2+r_-^2+r_+^2+ a_1^2+ a_2^2+L^2=0, \\ 
\label{crel2} &r_0^2 r_-^2 + r_0^2 r_+^2 + r_-^2 r_+^2 - a_1^2 a_2^2 + (2 M - a_1^2 - a_2^2)L^2  = 0  \\
\label{crel3} &r_0^2 r_-^2 r_+^2 + a_1^2 a_2^2L^2 = 0. 
\end{align}
Note that for the present five dimensional spacetime the mass parameter $M$ has a dimension of $(\text{length})^2$ and 
the Arnowitt-Deser-Misner mass and the angular momenta of this spacetime are specified as \cite{Gibbons:2004ai}
\be
{\cal{M}}=\df{\pi M (2\Xi_1+2\Xi_2-\Xi_1 \Xi_2)}{4\Xi_1^2 \Xi_2^2},\quad 
{\cal{J}}_1=\df{\pi M a_1}{2\Xi_1^2 \Xi_2},\quad {\cal{J}}_2=\df{\pi M a_2}{2\Xi_1  \Xi_2^2}.
\ee
Here, the two angular momenta ${\cal{J}}_1$ and ${\cal{J}}_2$ are in general different from each other.
The Hawking temperature and angular velocities at each horizon
$r_k\ (k=+,-,0)$ are given as
\be
T_k \coloneqq \df{r_k}{2\pi L^2}\df{(r_k^2-r_i^2)(r_k^2-r_j^2)}{(r_k^2+a_1^2)(r_k^2+a_2^2)},\quad \Omega_{k,1}\coloneqq  \df{a_1 \Xi_1}{r_k^2+a_1^2},\quad
\Omega_{k,2}\coloneqq \df{a_2 \Xi_2}{r_k^2+a_2^2},\quad i,j\neq k.
\ee
\subsection{Klein-Gordon equation}
\label{ssec:kgeq}

We consider the Klein-Gordon equation for a test massive scalar field in the Kerr-AdS$_5$ spacetime
\be (\Box-\mu^2)\Phi=0, \ee
where $\mu$ is the mass of the scalar field.
By separating the wave function as 
\be \Phi=R(r)S(\theta)e^{-i\omega t+im_1\phi_1+im_2\phi_2}, \ee 
we can separate the Klein-Gordon equation into angular and radial equations: 
\begin{align}
\label{KGang} &\Biggl[ 
\f{1}{\sin\theta\cos\theta}\f{d}{d\theta} \sin\theta\cos\theta \D_\theta \f{d}{d\theta} -\omega^2L^2 - \f{m_1^2\Xi_1}{\sin^2\theta} - \f{m_2^2\Xi_2}{\cos^2\theta} \notag\\
&~~ +\f{\Xi_1 \Xi_2}{\D_\theta}\mk{\omega L+m_1\df{a_1}{L}+m_2\df{a_2}{L}}^2 - \mu^2(a_1^2\cos^2\theta+a_2^2\sin^2\theta) + \sep \Biggr] S(\theta)=0 , \\
\label{KGrad} &\Biggl[ \f{1}{r}\f{d}{dr}r\D\f{d}{dr} + \f{(r^2+a_1^2)^2(r^2+a_2^2)^2}{r^4\D} \mk{ \omega - \f{m_1a_1\Xi_1}{r^2+a_1^2} - \f{m_2a_2\Xi_2}{r^2+a_2^2} }^2\notag\\
&~~ - \sep - \mu^2 r^2 - \f{1}{r^2} \{ a_1a_2\omega - a_2m_1\Xi_1 - a_1m_2\Xi_2 \}^2  \Biggr] R(r)=0 ,
\end{align}
where $\sep$ is a separation constant, or also called eigenvalue.

In the next subsection, we begin with the discussion of the angular function and the evaluation of the eigenvalue $\lambda$.
As we shall see below, the structures of the singularities of the angular equation~\eqref{KGang} are different for $|a_1|=|a_2|$ and $|a_1|\ne |a_2|$ cases.
Hence, we need a separate treatment for each case.

\subsection{Angular equation}
\label{ssec:ang}
Introducing a new coordinate and redefining the function, the angular equation can be transformed into a differential equation for a special function: hypergeometric function or Heun function~\cite{Aliev:2008yk, Amado:2017kao, BarraganAmado:2018zpa}. 
The eigenvalue $\lambda$ of the differential equation is determined by the regularity condition of those special functions. 
We shall formulate equal-/opposite-rotation case $a_1=\pm a_2$ in \S\ref{sssec:equal}, and different-rotation case $|a_1|\ne |a_2|$ in \S\ref{sssec:diff}.

\subsubsection{Equal-/opposite-rotation case}
\label{sssec:equal}

If $a_1=\pm a_2\eqqcolon a$, the angular equation~\eqref{KGang} can be solved in terms of hypergeometric function.
Applying the transformation
\begin{align} \label{ang_trans-sp}
x &= \sin^2\theta , \notag\\
S(\theta) &= x^{m_1/2}(x-1)^{m_2/2} v(x) , 
\end{align}
we obtain the hypergeometric equation
\be \label{anghyp}
\f{d^2v}{dx^2} + \mk{\f{c_a}{x}+\f{d_a}{x-1}} \f{dv}{dx} + \f{a_ab_a}{x(x-1)} v=0,
\ee
where 
\begin{align} 
a_a &= \f{1}{2} \kk{ m_1+m_2+1 + \sqrt{1+ \kk{\omega L+\f{a}{L}(m_1\pm m_2)}^2 + \f{\lambda-\omega^2L^2-\mu^2a^2}{\Xi}} } , \notag\\
b_a &= \f{1}{2} \kk{ m_1+m_2+1 - \sqrt{1+ \kk{\omega L+\f{a}{L}(m_1\pm m_2)}^2 + \f{\lambda-\omega^2L^2-\mu^2a^2}{\Xi}} } , \notag\\
c_a &= m_1+1, \quad d_a = m_2+1,
\end{align} 
and $\Xi=1-a^2/L^2$.
Note that $c_a+d_a = a_a+b_a +1$ holds.
The plus/minus sign corresponds to the each case of $a_1=\pm a_2=a$.
Note that, for the non-rotating case with $a=0$, the equation does not depend on $\omega$ and $\mu$.

The hypergeometric equation~\eqref{anghyp} possesses three regular singular points located at $x=0,1,\infty$.
Using the Frobenius method, we can construct linearly-independent two power series solutions at the vicinity of each regular singular point.
Let $v_{Ii}(x)$ denote such local solutions, 
where the subscript $I$ represents the regular singular point $I=0,1,\infty$ around which the local solution is defined, and $i=1,2$ is a label for the two independent local solutions at the regular singular point. 
The corresponding angular function $S_{Ii}(\theta)$ is defined via \eqref{ang_trans-sp}.
Among the four singular points,
we are interested in $x=0$ and $x=1$ since they correspond to $\theta=0,\pi$ and $\pi/2$, respectively.
Hence, for the following, we focus on the solution $v_{Ii}$ with $I=0,1$.

Around $x=0$ and $x=1$, we can construct two independent solutions, i.e., $v_{01}(x)$, $v_{02}(x)$ at $x=0$, and $v_{11}(x)$, $v_{12}(x)$ at $x=1$, which are given in terms of the hypergeometric function.
Specifically, two local solutions at $x=0$ are given by 
\begin{align}
\label{v01def} v_{01}(x) &= {}_2F_1(a_a, b_a, c_a;x), \\
\label{v02def} v_{02}(x) &= x^{1-c_a}{}_2F_1(a_a+1-c_a, b_a+1-c_a, 2-c_a;x),
\end{align}
and two local solutions at $x=1$ are given by
\begin{align}
\label{v11def} v_{11}(x) &= {}_2F_1(a_a, b_a, d_a;1-x), \\
\label{v12def} v_{12}(x) &= (1-x)^{1-d_a}{}_2F_1(c_a-a_a, c_a-b_a, 2-d_a;1-x),
\end{align}
where ${}_2F_1$ is the Gauss hypergeometric function.
The connection formula is given by
\begin{align} \label{conn}
v_{01}(x) &= \f{\Gamma(c_a)\Gamma(c_a-a_a-b_a)}{\Gamma(c_a-a_a)\Gamma(c_a-b_a)} v_{11}(x) + \f{\Gamma(c_a)\Gamma(a_a+b_a-c_a)}{\Gamma(a_a)\Gamma(b_a)} v_{12}(x),\\
v_{02}(x) &= \f{\Gamma(2-c_a)\Gamma(c_a-a_a-b_a)}{\Gamma(1-a_a)\Gamma(1-b_a)} v_{11}(x) + \f{\Gamma(2-c_a)\Gamma(a_a+b_a-c_a)}{\Gamma(a_a-c_a+1)\Gamma(b_a-c_a+1)} v_{12}(x).
\end{align}
At the vicinity of the regular singular points, the angular functions $S_{Ii} = x^{m_1/2}(x-1)^{m_2/2} v_{Ii}(x)$ behave as
\begin{align}
\notag
   & S_{01}(x) \propto x^{m_1 /2},\quad\quad\quad~\, 
   S_{02}(x)\propto x^{-m_1/2}, \quad\quad\quad~\,  (x\to 0), \\
   \label{yasympt}
   & S_{11}(x)\propto (1-x)^{m_2/2},\quad S_{12}(x)\propto (1-x)^{-m_2/2}, \quad  (x\to 1).
\end{align}

General solution for the angular function is given by a linear combination of the independent solutions.
Depending on the signs of $m_1,m_2$, we need to choose an appropriate pair of $S_{Ii}$ to maintain the regularity at $x=0,1$.
As an example, let us consider the case with $m_1\geq 0$ and $m_2\geq 0$.
In this case, $S_{01}(x)$ and $S_{11}(x)$ are regular at $x=0$ and $x=1$, respectively.
The two regularity conditions at both $x=0$ and $x=1$ are satisfied if $v_{01}(x)$ and $v_{11}(x)$ are linearly dependent.
From the connection formula~\eqref{conn}, we see that they are linearly dependent if the coefficient of $v_{12}(x)$ vanishes, i.e.,
\be a_a=-j, ~~~\text{or}~~~  b_a=-j, \ee
where $j$ is a non-negative integer.
This condition singles out special discrete values of $\lambda$.
By solving the above relation, we can write down the eigenvalue $\lambda$ as
\be
\lambda=\Xi \left[ (2j+m_1+m_2)(2j+m_1+m_2+2)-2\omega a (m_1\pm m_2)-\dfrac{a^2}{L^2}(m_1\pm m_2)^2 \right]+a^2(\omega^2+\mu^2).
\ee
Recall that the plus/minus sign corresponds to the each case of $a_1=\pm a_2=a$.
This expression recovers the formula obtained in \cite{Aliev:2008yk} for the case $a_1=a_2=a$ and $m_1\geq 0$ and $m_2\geq 0$.

Similarly, for the other three cases, $m_1\geq 0$ and $m_2< 0$, $m_1< 0$ and $m_2\geq 0$, and $m_1< 0$ and $m_2< 0$, we can identify the regularity condition from the pole of the gamma functions in the coefficient of singular solution for each case.
As a result, with $a_1=\pm a_2=a$ and arbitrary $m_1,m_2$, the angular function is regular at both singular points, i.e., $S(x) \propto x^{|m_1| /2}$ at $x=0$ and $S(x)\propto (1-x)^{|m_2|/2}$ at $x=1$, if the eigenvalue takes the value 
\begin{align} \label{lam-eqrot}
\lambda&=\Xi \left[\ell(\ell+2)-2\omega a (m_1\pm m_2)-\dfrac{a^2}{L^2}(m_1\pm m_2)^2 \right]+a^2(\omega^2+\mu^2), \notag\\ 
\ell &\equiv 2j+|m_1|+|m_2|,
\end{align}
where $j$ is a non-negative integer.
Note that the definition of $\ell$ is generalized from the one used in \cite{Aliev:2008yk}.
Here, $\ell$ and $m_1,m_2$ play a role of the orbital quantum number, and magnetic quantum numbers, respectively. 
For the non-rotating case with $a=0$, the eigenvalue reduces to $\lambda=\ell(\ell+2)$.
Note that $\ell \geq |m_1|+|m_2|$ holds, and hence $\ell$ takes even/odd values if the sum $|m_1|+|m_2|$ is even/odd. 
Specifically, possible combinations for lower $\ell$ are:
\begin{align} \label{ellm1m2}
\ell&=0: ~~(m_1,m_2)=(0,0), \notag\\
\ell&=1: ~~(m_1,m_2)=(0,\pm 1),~ (\pm 1,0), \notag\\
\ell&=2: ~~(m_1,m_2)=(0,0),~ (\pm 1,\pm 1),~ (0,\pm 2),~ (\pm 2,0),\notag\\
\ell&=3: ~~(m_1,m_2)=(0,\pm 1),~ (\pm 1,0),~ (\pm 1,\pm 2),~ (\pm 2,\pm 1),~ (0,\pm 3),~ (\pm 3,0),
\end{align}
where any double signs are allowed, which should not be conflated with the sign corresponding to $a_1=\pm a_2$.
For the above counting classified by the value of $\ell$, one should be careful to exhaust all possible cases for $j$.
For instance, for $\ell=2$, $(m_1,m_2)=(0,0)$ originates from $j=1$, whereas $(m_1,m_2)=(\pm 1,\pm 1), (0,\pm 2), (\pm 2,0)$ come from $j=0$.
Note also that combinations such as $(\ell,m_1,m_2)=(2,\pm 1,0), (3,\pm 2,0), (3,\pm 1,\pm 1)$ are not allowed.

\subsubsection{Different-rotation case}
\label{sssec:diff}
If $|a_1|\ne |a_2|$,
the equation for the angular function is transformed into the Heun equation. 
For the angular equation~\eqref{KGang}, we apply the transformation
\begin{align} \label{ang_trans}
x &= \sin^2\theta , \notag\\
S(\theta) &= x^{m_1/2}(x-1)^{m_2/2}(x-x_0)^{\tau/2} w(x) , 
\end{align}
with
\be 
\label{x0tau}
x_0=\f{\Xi_1}{\Xi_1-\Xi_2}, \quad 
\tau=\omega L+m_1\df{a_1}{L}+m_2\df{a_2}{L}, \ee
to transform it into the Heun equation
\be \label{angHeun} \f{d^2w}{dx^2} + \mk{\f{\gamma_a}{x} + \f{\delta_a}{x-1} + \f{\epsilon_a}{x-x_0}} \f{dw}{dx} 
+ \f{\alpha_a\beta_a x-q_a}{x(x-1)(x-x_0)} w
= 0 ,  \ee
where parameters are defined as
\begin{align} 
q_a&= \dfrac{1}{2}\left\{(\tau+m_1)(m_1+1)+x_0(m_1+m_2+m_1 m_2)-\dfrac{x_0}{2\Xi_1}[\lambda -\mu^2 a_1^2 -\omega^2L^2+\Xi_2(\tau^2-m_1^2-m_2^2 ) ]\right\},  \notag\\
\alpha_a&= \f{1}{2} ( m_1 + m_2 + \tau + 2 + \sqrt{\mu^2L^2+4} ), \quad 
\beta_a= \f{1}{2} ( m_1 + m_2 + \tau + 2 - \sqrt{\mu^2 L^2+4} ),\notag\\ 
\gamma_a&= m_1+1, \quad 
\delta_a=m_2+1, \quad 
\epsilon_a=\tau+1,
\end{align}
and they satisfy
\begin{equation}
    \gamma_a+\delta_a+\epsilon_a =\alpha_a +\beta_a +1, \quad\quad x_0\neq 0,1,\infty.
\end{equation}
Note that $x_0$ becomes singular if $a_1=\pm a_2$.  
In this case we need a separate treatment presented in \S\ref{sssec:equal}.

The Heun equation~\eqref{angHeun} possesses four regular singular points located at $x=0,1,x_0,\infty$.
Along the same lines as the equal-/opposite-rotation case, we can construct power series solution at the vicinity of each regular singular point, which are denoted by $w_{Ii}(x)$, and can be written down in terms of local Heun function.
Namely, two local solutions at $x=0$ are given by
\begin{align}
\label{w01def} w_{01}(x) &= Hl(x_0, q_a;\alpha_a, \beta_a, \gamma_a, \delta_a;x), \\
\label{w02def} w_{02}(x) &= x^{1-\gamma_a}Hl(x_0, (x_0\delta_a+\epsilon_a)(1-\gamma_a)+q_a;\alpha_a+1-\gamma_a, \beta_a+1-\gamma_a, 2-\gamma_a, \delta_a;x),
\end{align}
and two local solutions at $x=1$ are given by
\begin{align}
\label{w11def} w_{11}(x) &= Hl(1-x_0, \alpha_a\beta_a-q_a;\alpha_a, \beta_a, \delta_a, \gamma_a;1-x), \\
\label{w12def} w_{12}(x) &= (1-x)^{1-\delta_a}Hl(1-x_0, ((1-x_0)\gamma_a+\epsilon_a)(1-\delta_a)+\alpha_a\beta_a-q_a;\alpha_a+1-\delta_a, \beta_a+1-\delta_a, 2-\delta_a, \gamma_a;1-x),
\end{align}
where $Hl$ denotes the local Heun function~\cite{Heun1889,ronveaux1995heun,slavianov2000special,Maier_2006}.

The asymptotic behaviors of these solutions are as follows:
\begin{align}
\notag
   & w_{01}(x)=1+{\cal{O}}(x),\quad\quad~~ w_{02}(x)=x^{1-\gamma_a}[1+{\cal{O}}(x)],\quad\quad\quad\quad\quad\,  (x\to 0),  \\
   \label{yasympt}
   & w_{11}(x)=1+{\cal{O}}(1-x),\quad  w_{12}(x)=(1-x)^{1-\delta_a}[1+{\cal{O}}(1-x)],\quad  (x\to 1), 
\end{align}
Therefore, the behavior of the angular function $S_{Ii} = x^{m_1/2}(x-1)^{m_2/2} w_{Ii}(x)$ at the vicinity of the singular point is the same as \eqref{yasympt}.
For the angular function to satisfy the regularity at the singularity point, we 
choose $S_{01}$ and $S_{11}$ for positive magnetic quantum numbers and $S_{02}$ and $S_{12}$ for negative ones. 
To satisfy both regularity conditions at $x=0$ and $x=1$,
the linear dependence of the solutions is required, which is represented by the following condition:
\be \label{ang_reg}
W_x[w_{0i},w_{1j}]=0,\quad
i=\begin{cases}
1, & (m_1\geq 0) , \\
2, & (m_1< 0) ,
\end{cases} \quad
j=\begin{cases}
1, & (m_2\geq 0) , \\
2, & (m_2< 0) ,
\end{cases} 
\ee
where 
$W_x[f,g] = f \f{dg}{dx} - \f{df}{dx} g$
is a Wronskian. 
The above condition determines the eigenvalue $\lambda$ in \eqref{KGang} and \eqref{KGrad} for the different-rotation case $a_1\ne \pm a_2$.

For practical computation, we shall obtain the eigenvalue~$\lambda$ by applying a root-finding algorithm to \eqref{ang_reg}, and hence we need to input an initial value of $\lambda$ for the algorithm.
As the initial value, we can adopt the eigenvalue for the equal-/opposite-rotation case given in \eqref{lam-eqrot}.
Note that, while the expression~\eqref{lam-eqrot} provides the eigenvalue $\lambda$ for the equal-rotation case,
we can use it as a sufficiently good initial value for the root-finding algorithm for the different-rotation case so long as the difference between the spin parameters is not so large.
If the difference between the spin parameters is large, we can begin with obtaining $\lambda$ for the equal-/opposite-rotation case and iterate the root-finding process by gradually increasing or decreasing the spin parameter until it reaches the value of interest.

\subsection{Radial equation}
\label{ssec:rad}

The radial equation~\eqref{KGrad} has four regular singular points both for 
equal-/opposite-rotation and different-rotation cases. Regarding $r^2$ as an independent 
variable, the four regular singular points are located at $r^2=r_0^2,r_+^2,r_-^2, \infty$.
By applying the transformation
\begin{align} \label{rad_trans}
z &= \f{r^2-r_+^2}{r^2-r_0^2},\notag\\ 
R &= z^{\theta_+/2}(z-1)^{\sigma/2}(z-z_0)^{\theta_-/2} y(z) , 
\end{align}
where 
\be z_0=\f{r_-^2-r_+^2}{r_-^2-r_0^2}, \quad 
\sigma=2+\sqrt{4+\mu^2L^2}, \quad 
\theta_k=\f{i(\omega-m_1\Omega_{k,1}-m_2\Omega_{k,2})}{2\pi T_k},  
\ee
we transform the radial equation into the Heun equation
\be \label{radHeun} \f{d^2y}{dz^2} + \mk{\f{\gamma_r}{z} + \f{\delta_r}{z-1} + \f{\epsilon_r}{z-z_0}} \f{dy}{dz} + \f{\alpha_r\beta_r z-q_r}{z(z-1)(z-z_0)} y = 0 , \ee
where
\begin{align} 
q_r&= \f{1}{4} \Biggl[ \f{\omega^2L^2-\sep-r_+^2\mu^2}{r_-^2-r_0^2}\ L^2 +2z_0(\sigma \theta_+  + \sigma - \theta_+ ) + (\theta_+ + \theta_-)(\theta_+ + \theta_- + 2) - \theta_0^2 \Biggr], \notag\\
\alpha_r&= \f{1}{2}(\sigma + \theta_ + + \theta_- + \theta_0), \quad 
\beta_r= \f{1}{2}(\sigma + \theta_+ + \theta_- - \theta_0), \notag\\ 
\gamma_r&= 1+\theta_+, \quad 
\delta_r=-1+\sigma, \quad 
\epsilon_r=1+\theta_-.
\end{align}
Note that $\gamma_r+\delta_r+\epsilon_r=\alpha_r+\beta_r+1$ holds.
After the transformation, the four regular singular points $r^2=r_0^2,r_+^2,r_-^2, \infty$ are mapped to $z=\infty,0,z_0,1$, respectively.
For the AdS case we are interested in the scattering problem at the region $r_+<r<\infty$, which is now $0<z<1$ for the new variable.

The sets of the linearly independent local solutions are given as 
\begin{align}
\label{y01def} y_{01}(z) &= Hl(z_0, q_r;\alpha_r, \beta_r, \gamma_r, \delta_r;z), \\
\label{y02def} y_{02}(z) &= z^{1-\gamma_r}Hl(z_0, (z_0\delta_r+\epsilon_r)(1-\gamma_r)+q_r;\alpha_r+1-\gamma_r, \beta_r+1-\gamma_r, 2-\gamma_r, \delta_r;z),
\end{align}
around $z=0$ and  
\begin{align}
\label{y11def} y_{11}(z) &= Hl(1-z_0, \alpha\beta-q_r;\alpha_r, \beta_r, \delta_r, \gamma_r;1-z), \\
\label{y12def} y_{12}(z) &= (1-z)^{1-\delta_r}Hl(1-z_0, ((1-z_0)\gamma_r+\epsilon_r)(1-\delta_r)+\alpha_r\beta_r-q_r;\alpha_r+1-\delta_r, \beta_r+1-\delta_r, 2-\delta_r, \gamma_r;1-z).
\end{align}
around $z=1$, respectively.

A caveat is that the above formulation does not work for exactly massless case.
For the massless case, since $\delta_r=3$, the characteristic exponents around $z=1$, which are $0$ and $1-\delta_r$, are separated by an integer.
In this case, the local Heun function used in \eqref{y12def} does not exist.
We should thus modify the definition of $y_{12}(z)$ to another independent solution by using a logarithmic function.
While our formulation would also work for massless case after this modification, for the following, we do not discuss this case for simplicity, and focus on the massive case $\mu\ne 0$. 
As another caveat, the method with the Heun function cannot be used in the exact extremal case since 
the number of the regular singular points will be changed, but we can compute near extremal case unless $\Delta=0$ has a double root at the outer horizon.

\subsection{Connection coefficients and asymptotic behavior}
\label{sec:asymp}
In this subsection, we discuss the asymptotic behavior of the solutions of the radial equation \eqref{KGrad} in terms of the local Heun functions around $z=0$ and $z=1$ given in \eqref{y01def}--\eqref{y12def}. 
First, the asymptotic behaviors of these solutions are as follows:
\begin{align}
\notag
   & y_{01}(z)=1+{\cal{O}}(z),\quad~~~~~ y_{02}(z)=z^{1-\gamma_r}[1+{\cal{O}}(z)],\\
   \label{yasympt}
   & y_{11}(z)=1+{\cal{O}}(1-z),\quad y_{12}(z)=(1-z)^{1-\delta_r}[1+{\cal{O}}(1-z)],
\end{align}
which shall be used when we identify the in/outgoing waves below.

The local Heun functions at $z=0$ are related to the local Heun functions at $z=1$ via linear combinations
\begin{align}
y_{01}(z)&= C_{11} y_{11}(z)+C_{12}y_{12}(z) , \label{y01rel} \\
y_{02}(z)&= C_{21}y_{11}(z)+C_{22}y_{12}(z) . \label{y02rel} 
\end{align}
The connection coefficients can be obtained as a ratio of the Wronskians as
\be \label{C11} C_{11}=\f{W_z[y_{01}, y_{12}]}{W_z[y_{11}, y_{12}]}, \quad
C_{12}=\f{W_z[y_{01}, y_{11}]}{W_z[y_{12}, y_{11}]}, \quad 
C_{21}=\f{W_z[y_{02}, y_{12}]}{W_z[y_{11}, y_{12}]}, \quad 
C_{22}=\f{W_z[y_{02}, y_{11}]}{W_z[y_{12}, y_{11}]},
\ee
where $W_z[u,v] = u \f{dv}{dz} - \f{du}{dz} v$.
Note that from \eqref{KGrad} it holds that, for linearly independent solutions $y_a,y_b$, 
\be z^{\gamma_r}(z-1)^{\delta_r}(z-z_r)^{\epsilon_r}W_z[y_a,y_b] = {\rm const}. \ee
Therefore, while the Wronskian itself is not constant, the ratio between two Wronskians is constant.

Conversely, the local Heun functions at $z=1$ can be expressed as
\begin{align}
y_{11}(z) = D_{11}y_{01}(z)+D_{12}y_{02}(z), \label{y11rel} \\ 
y_{12}(z) = D_{21}y_{01}(z)+D_{22}y_{02}(z), \label{y12rel}  
\end{align}
where 
\be 
\begin{pmatrix}
D_{11} & D_{12} \\
D_{21} & D_{22}
\end{pmatrix}
=
\begin{pmatrix}
C_{11} & C_{12} \\
C_{21} & C_{22}
\end{pmatrix}^{-1}
=
\f{W_z[y_{11}, y_{12}]}{W_z[y_{01}, y_{02}]}
\begin{pmatrix} \label{CDrel}
C_{22} & -C_{12} \\
-C_{21} & C_{11}
\end{pmatrix}
\ee
namely,
\be \label{D11} D_{11}=\f{W_z[y_{11},y_{02}]}{W_z[y_{01},y_{02}]}, \quad
D_{12}=\f{W_z[y_{11},y_{01}]}{W_z[y_{02},y_{01}]}, \quad 
D_{21}=\f{W_z[y_{12},y_{02}]}{W_z[y_{01},y_{02}]}, \quad
D_{22}=\f{W_z[y_{12},y_{01}]}{W_z[y_{02},y_{01}]}.\ee

Considering a boundary condition suitable for problems to solve such as wave scattering, Hawking radiation, or QN modes, one can compute the reflection amplitude, greybody factor, or QN frequencies with the above connection coefficients between local Heun functions.
Recalling \eqref{rad_trans}, let us define
\begin{equation}
    R_{Ii}=z^{\theta_+/2}(z-1)^{\sigma/2}(z-z_0)^{\theta_-/2} y_{Ii}(z),
\end{equation}
with $I=0,1$ and $i=1,2$.
To identify whether each mode $R_{Ii}$ corresponds to the in/outgoing wave at an asymptotic region, we examine the asymptotic 
behavior of the tortoise coordinate, which is defined by 
\be
\label{tort}
\df{dr_*}{dr}=\df{(r^2+a_1^2)(r^2+a_2^2)}{r^2\D},
\ee
and we obtain 
\be \label{rstar} r_*= \f{1}{4\pi T_0} \ln \left|\f{r-r_0}{r+r_0}\right| + \f{1}{4\pi T_+} \ln \left|\f{r-r_+}{r+r_+}\right| + \f{1}{4\pi T_-} \ln \left|\f{r-r_-}{r+r_-}\right| . \ee
Note that the region $r_+<r<\infty$ of our interest amounts to $-\infty<r_*<0$.
At the vicinity of each regular singular point, the 
tortoise coordinate behaves as
\be
r_* \rightarrow \df{1}{4\pi T_k}\ln \left|\f{r-r_k}{2r_k}\right|+c_k \quad (r \rightarrow r_k ),\qquad
 c_k\coloneqq 
 \sum_{j\ne k}\df{1}{4\pi T_j}\ln \left|\f{r_k-r_j}{r_k+r_j}\right|. \ee
Conversely, we obtain
\begin{equation} \label{rrstar} 
    r-r_k \simeq 2r_k e^{-4\pi c_k T_k } e^{4\pi T_k r_*} \quad (r \rightarrow r_k ).
\end{equation}

Using \eqref{yasympt} and \eqref{rrstar}, 
the asymptotic behavior of the radial function at $r\to r_+$ $(z \to 0)$ are given by 
\begin{align}
\label{R01asym}
R_{01}& \simeq A_{01}e^{i\tilde{\omega}r_*}, \qquad\,\,\, 
A_{01}\coloneqq \mk{\df{4r_+^2 e^{-4\pi c_+ T_+} }{r_+^2-r_0^2}}^{\theta_+/2}(-z_0)^{\theta_-/2}(-1)^{\sigma/2}, \\
\label{R02asym}
R_{02} &\simeq A_{02}e^{-i\tilde{\omega}r_*}, \qquad 
A_{02}\coloneqq \mk{\df{4r_+^2 e^{-4\pi c_+ T_+} }{r_+^2-r_0^2}}^{-\theta_+/2}(-z_0)^{\theta_-/2}(-1)^{\sigma/2},
\end{align}
where $\tilde{\omega}\coloneqq \omega-m_1 \Omega_{+,1}-m_2 \Omega_{+,2}$.
Therefore, $R_{01}$ and $R_{02}$ correspond to the outgoing and ingoing modes, respectively. 
On the other hand, the asymptotic behavior of the radial function at $r\rightarrow  \infty$ $(z \simeq 1)$ are given by
\begin{align}
\label{R11asym}
R_{11} &\simeq A_{11}r^{-\sigma}, \qquad~
A_{11}\coloneqq (1-z_0)^{\theta_-/2} \mk{r_0^2-r_+^2}^{\sigma/2}, \\
\label{R12asym}
R_{12} &\simeq A_{12}r^{\sigma-4}, \qquad 
A_{12}\coloneqq (1-z_0)^{\theta_-/2}(-1)^{\sigma/2}\mk{r_+^2-r_0^2}^{2-\sigma/2}.
\end{align}
Since $\sigma\geq 4$, $R_{11}$ and $R_{12}$ are the decaying and growing modes, respectively.

Using the asymptotic behaviors of the radial functions with proper boundary conditions, it is 
possible to compute the quantities such as the QN frequency and the greybody factor. Note that in our formalism, we do not consider any approximation, hence it can be utilized for parameters in 
a wide range.


Before discussing the QN modes and the greybody factor, let us explicitly write down coefficients for the asymptotic form of the radial function, i.e., 
\be R(r) \rightarrow {B_1}\ r^{-\sigma}+{B_2}\ r^{\sigma-4},\quad (r\rightarrow \infty). \ee
The coefficients $B_1$ and $B_2$ connect the local solution around the event horizon, which we assume satisfies a certain boundary condition, and the local solution at the infinity.
The asymptotic form and the coefficients play an important role in the context of the gauge/gravity correspondence \cite{Natsuume:2014sfa}.
For instance, let us impose the purely ingoing boundary condition at the black hole horizon, $R(r) \rightarrow e^{-i\tilde{\omega}r_*}$ for $r\to r_+$.
The solution satisfying the boundary condition can be written as $R(r)=R_{02}(r)/A_{02}$.
Taking into account of the definitions of $y(z)$ \eqref{rad_trans} and the connection coefficients \eqref{y02rel}, and using the asymptotic forms  \eqref{R11asym} and \eqref{R12asym}, 
the coefficients $B_1$ and $B_2$ can be written as
\begin{align}
\notag B_1&=\df{C_{21}A_{11}}{A_{02}}=(-1)^{-\sigma/2} \f{W_z[y_{02}, y_{12}]}{W_z[y_{11}, y_{12}]}\left(\df{z_0-1}{z_0}\right)^{\theta_-/2}\mk{\df{4r_+^2 e^{-4\pi c_+ T_+} }{r_+^2-r_0^2}}^{\theta_+/2}\mk{r_0^2-r_+^2}^{\sigma/2},\\
B_2&=\df{C_{22}A_{12}}{A_{02}}=  \f{W_z[y_{02}, y_{11}]}{W_z[y_{12}, y_{11}]}\left(\df{z_0-1}{z_0}\right)^{\theta_-/2}\mk{\df{4r_+^2 e^{-4\pi c_+ T_+} }{r_+^2-r_0^2}}^{\theta_+/2}(r_+^2-r_0^2)^{2-\sigma/2} .
\label{C0_C1}
\end{align}
It is also straightforward to obtain analytic expressions of coefficients corresponding to the purely outgoing boundary condition at the black hole horizon.
In the context of the gauge/gravity correspondence, these coefficients are basically computed numerically in the literature (see e.g., \cite{Hartnoll:2009sz}).
A holographic interpretation is given based on the GKP-Witten relation \cite{Gubser1998,Witten1998}, which represents the duality between the partition function of the quantum field theory on the AdS boundary and the on-shell action of the bulk classical gravity.
Although a specific application is beyond the scope of the present paper, the above analytic expressions in terms of the exact solution of the Klein-Gordon equation may be useful for investigation of dual phenomena.

\section{Quasinormal Modes}
\label{sec:qnm}
The QN modes are damped oscillations with a series of discrete frequencies that satisfy a certain set of boundary conditions.
Here, we require the purely ingoing boundary 
condition at the black hole horizon and the Dirichlet or decaying boundary condition at the conformal infinity.
In our formalism, this boundary condition corresponds to setting the coefficients of the growing mode to be zero in the following IN mode:
\begin{align} 
\label{Rin-Heun} R_{{\rm in}}(r) &= \begin{cases}
R_{02}(r), & (r\to r_+) ,\\
C_{21}R_{11,s}(r)+C_{22}R_{12}(r), & (r \gg r_+).
\end{cases}\\
&\rightarrow \begin{cases}
A_{02}e^{-i\tilde{\omega}r_*}, & (r\to r_+) ,\\
C_{21}A_{11}r^{-\sigma}+C_{22}A_{12}r^{\sigma-4}, & (r \gg r_+).
\end{cases}
\end{align} 
As $A_{12}\neq 0$ generally, we search frequencies satisfying $C_{22}=0$ to obtain the 
QN mode frequencies.
This strategy is also adopted recently in \cite{Koga:2022vun} to study the QN modes of the Kerr-AdS$_5$ with $m_1,m_2\geq 0$. 
However, as discussed in \S\ref{ssec:ang}, negative $m_1$ and $m_2$ are also allowed.
Below, we consider the QN modes with all possible magnetic quantum numbers.

\begin{figure}[H]
 \centering
 \includegraphics[width=0.6\linewidth]{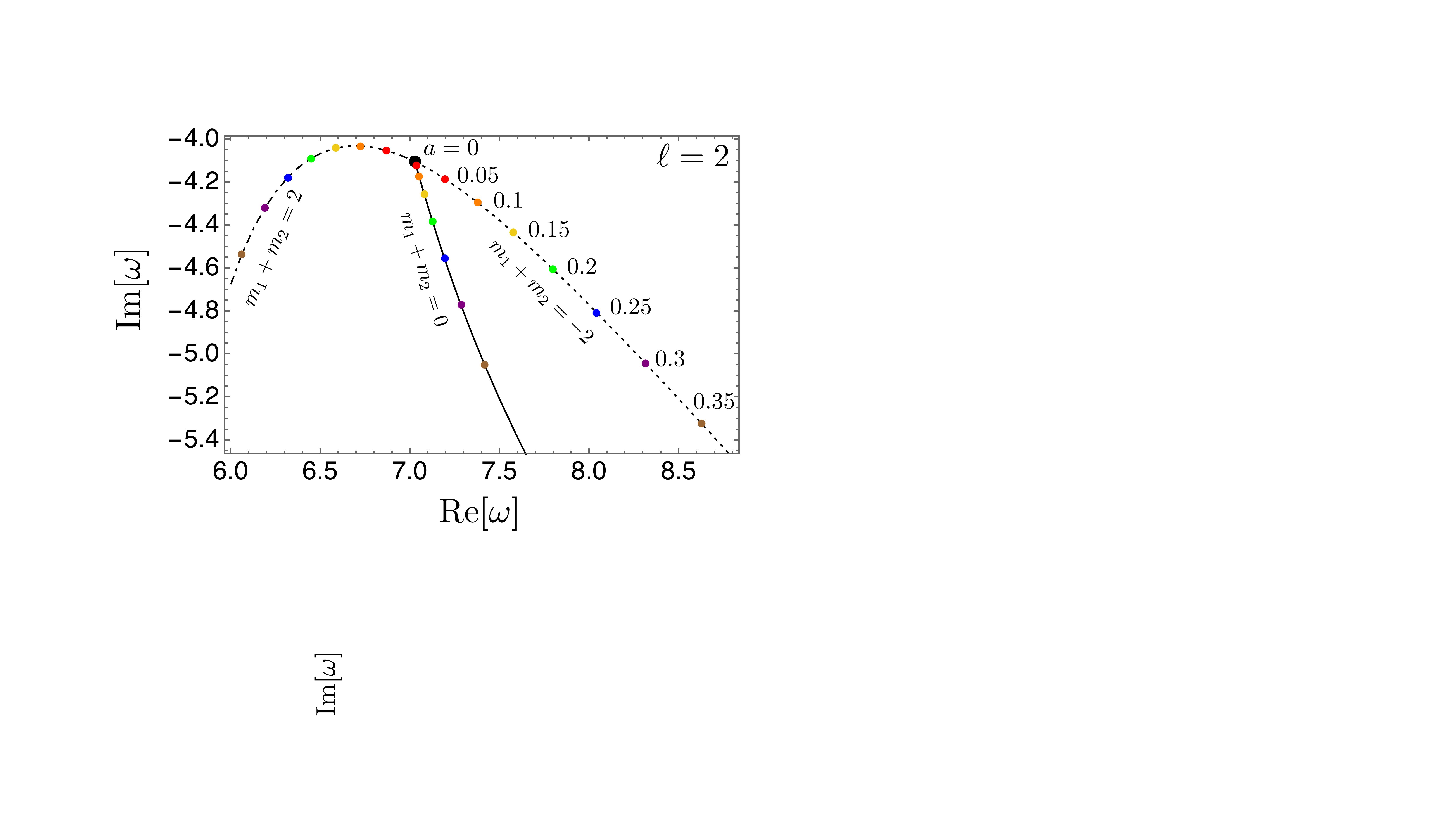}
 \caption{\footnotesize{QN mode frequencies of $\ell=2$ modes for equal-rotation case $a_1=a_2=a$ with $M=5, L=1, \mu=0.01$. 
 The large black dot is QN mode frequency for nonrotating case with $a=0$.
 For $a\neq0$, the spectrum splits into three series depending on the value of $m_1+m_2$. 
 Solid, dashed, dot-dashed curves correspond to the QNMs with $m_1+m_2=0$, $-2$, $2$, respectively.
 }}
 \label{fig:QNMeq}
\end{figure}

\begin{figure}[H]
 \centering
 \includegraphics[width=0.6\linewidth]{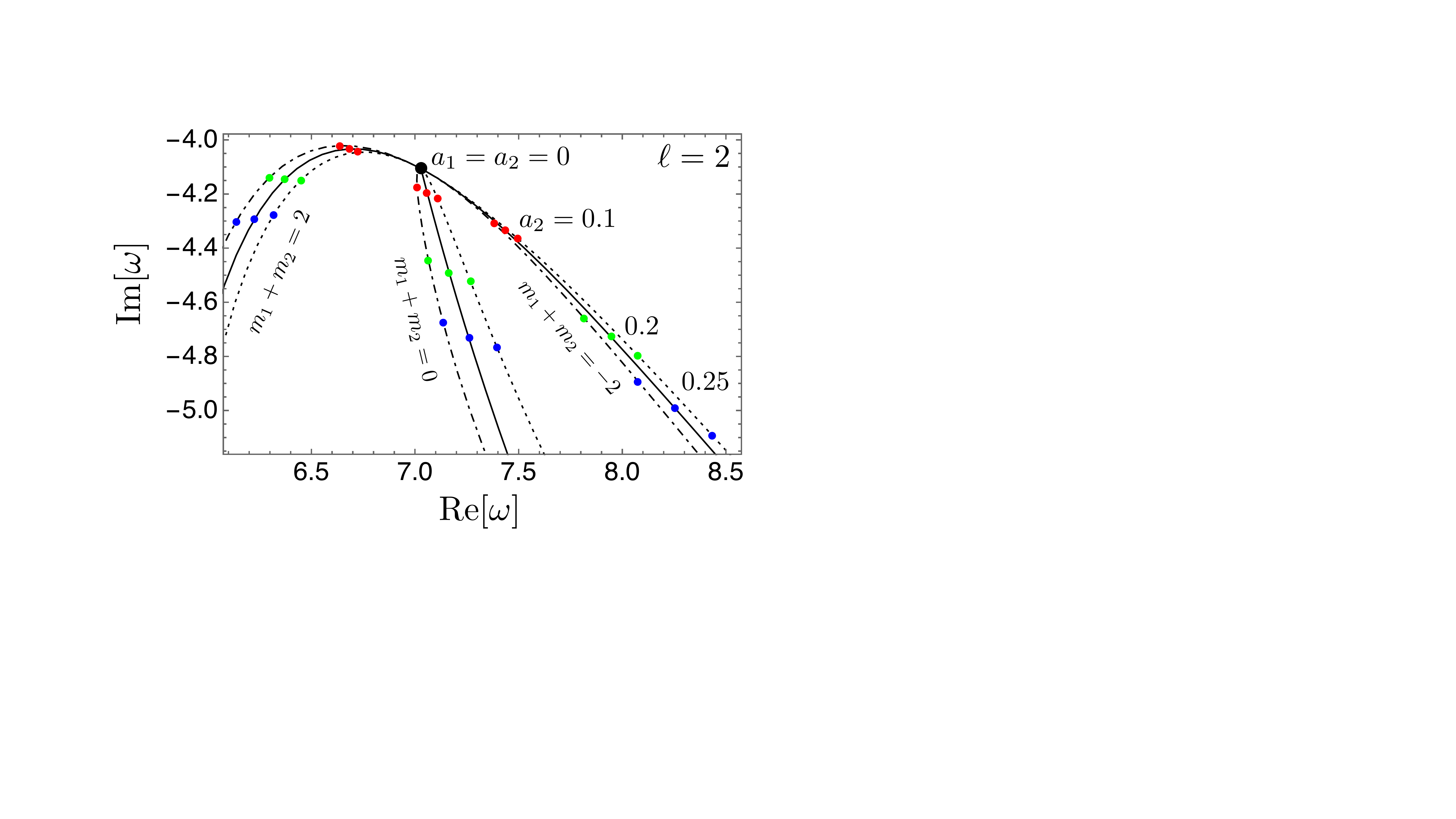}
 \caption{\footnotesize{QN mode frequencies 
 of $\ell=2$ modes for different-rotation case ($a_1=1.3 a_2$) with $M=5, L=1, \mu=0.01$. 
 The large black dot is QN mode frequency for $a_1=a_2=0$.
 Compared to the equal-rotation case in Fig.~\ref{fig:QNMeq}, there are extra splitting due to 
 the multi-valued angular momenta.}}
 \label{fig:QNMdif}
\end{figure}

In Figs.~\ref{fig:QNMeq} and \ref{fig:QNMdif}, 
we show QN modes for relatively small spin parameters.
We consider all possible magnetic quantum numbers for $\ell=2$ listed in \eqref{ellm1m2}, 
including negative $m_1$ and $m_2$.
The degeneracy between QN modes for nonrotating black hole is broken by nonvanishing spin parameter. 
This Zeeman-like splitting is well-known in the QN modes for the four-dimensional Kerr black hole~\cite{Onozawa:1996ux,Berti:2009kk}, and now confirmed for the ones for the Kerr-AdS$_5$. 
As shown in Fig.~\ref{fig:QNMeq}, for the equal-rotation case $a_1=a_2=a$, there are three 
series of QN modes depending on the value of the total magnetic quantum number $m_1+m_2$.
The same set of curves is obtained for the opposite-rotation case since the QN modes basically depend 
on the combinations $m_1a_1$ and $m_2a_2$.

In Fig.~\ref{fig:QNMdif}, we consider the case with different angular momenta $a_1 \neq \pm a_2$.
We fix the ratio as $a_1 =1.3a_2$, and float $a_2$ as $0.1, 0.2$, and $0.25$. 
In this case, each of the three curves in Fig.~\ref{fig:QNMeq} for the equal-/opposite-rotation case 
splits into three curves even for the same value of $m_1+m_2$. 
For example, the case $m_1+m_2=0$ includes three cases $(m_1,m_2)=(0,0),\ (1,-1),\ (-1,1)$, which are degenerate for the equal-/opposite-rotation case but distinct for the different angular momenta case.
As a result, there 
exist nine series of QN modes in total.
Thus, the degeneracy of the QN modes is two-fold structure: The first degeneracy is nonrotation vs rotation, which also exists in the four-dimensional Kerr case, and the second degeneracy is equal-/opposite-rotation vs different-rotation, which is peculiar to the higher-dimensional black holes.

There exists another series of QN modes called 
type II in \cite{Koga:2022vun}, of which real parts 
are localized near the threshold frequency for superradiance: 
$\text{Re}[\omega_\text{QNM}]\sim m_1 \Omega_{+,1}+m_2 \Omega_{+,2}$.
For this sequence, the absolute values of the imaginary part of the QN frequencies become smaller as the spin parameter gets closer in the extremal limit.
The real parts of the QN frequencies approach zero in the extremal limit as $m_1 \Omega_{+,1}+m_2 \Omega_{+,2}\to 0$ in this limit.
For lower spin values, these modes move away from the imaginary axis.

\begin{figure}[H]
 \centering
 \includegraphics[width=0.6\linewidth]
 {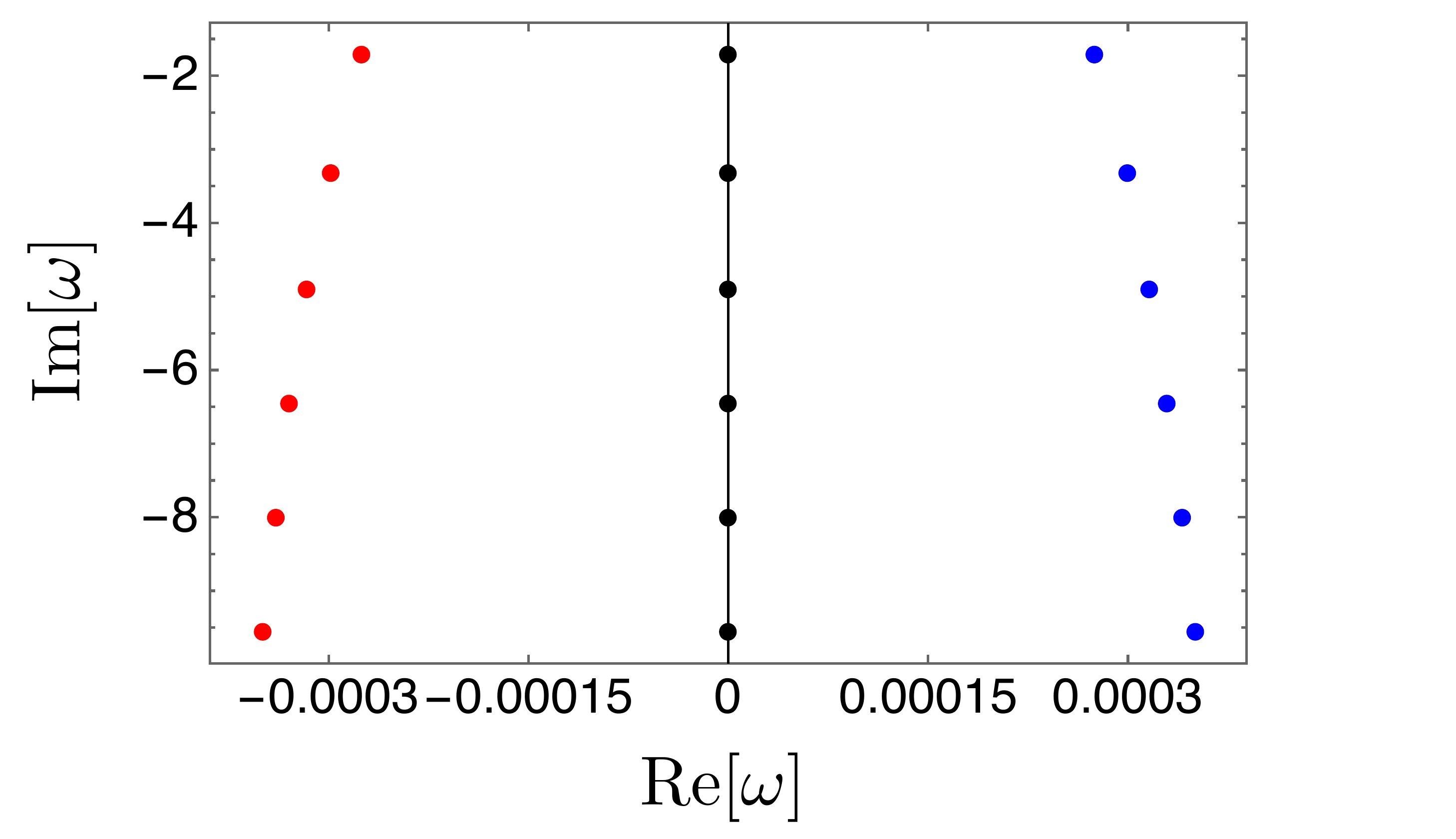}
 \caption{\footnotesize{Overtones of the QN mode frequencies of $\ell=2$ modes for $a_1=a_2=0.9999$ with $M=5, L=1, \mu=0.01$. The coloring is given as 
 $m_1+m_2=-2$ (red), $m_1+m_2=0$ (black), $m_1+m_2+2$ (blue)}.}
 \label{fig:a09999_overtone}
\end{figure}

Furthermore, we find that there exists another sequence of purely imaginary modes corresponding to $m_1+m_2=0$.
In Fig.~\ref{fig:a09999_overtone}, we present fundamental modes and overtones for these sequences 
for the case of near extremal equal spin $a_1=a_2=0.9999$.
The two sequences on the both sides of the imaginary axis match the modes studied in \cite{Koga:2022vun}.
We find the purely imaginary modes which are almost at the middle points of the above modes.
Their real parts are almost vanishing within the numerical error. 
As we shall see below, the real parts of these modes are still almost vanishing even for lower spin values.

Note that the purely imaginary modes satisfy the properties of the type II QN modes, 
and therefore may belong to the type II QN modes.
Indeed, the real part of the type II QN modes, $m_1\Omega_{+,1}+m_2\Omega_{+,2}$, should be vanishing for the equal-rotation case with $m_1+m_2=0$, which is consistent with the purely imaginary modes found here.
The purely imaginary modes have almost 
equal spacing, and 
the interval almost coincides with $2\pi T_+$. 
As an example, the value of the imaginary part of fifth and fourth overtones in Fig.~\ref{fig:a09999_overtone} 
are $-9.55255$ and $-8.00414$, respectively, 
and hence the interval is $1.5481$, 
which is indeed close to $2\pi T_+ = 1.5487$. 
For higher overtones the interval gets closer to $2\pi T_+$.
The equidistant property for the imaginary parts is known for four-dimensional near extremal Kerr spacetime as derived with analytic computation by Hod \cite{Hod:2013fea}, and also is observed for the type II QN modes in the Kerr-AdS$_5$ spacetime~\cite{Koga:2022vun}. 

As mentioned in \S\ref{sec:intro}, the purely imaginary modes also show up for the Schwarzschild, Kerr, and other spacetimes and play an important role. 
It is known that such modes have peculiar properties~\cite{MaassenvandenBrink:2000iwh,Cook:2016fge,Cook:2016ngj} (see also Appendix~A in \cite{Berti:2009kk}).
For instance, the nature of the boundary conditions at the Schwarzschild algebraically special frequency is known to be extremely subtle, and such modes actually consist of the QN modes and so-called total transmission modes.
At this moment, the physical meaning of the purely imaginary modes for the scalar field in the Kerr-AdS$_5$ spacetime is not clear. 
They may or (partially) may not be the QN modes.
Hence, we call them just purely imaginary modes, though we may call them QN modes sometimes for simplicity when we treat them together with other series of QN modes.

\begin{figure}[H]
 \centering
 \includegraphics[width=0.9\linewidth]{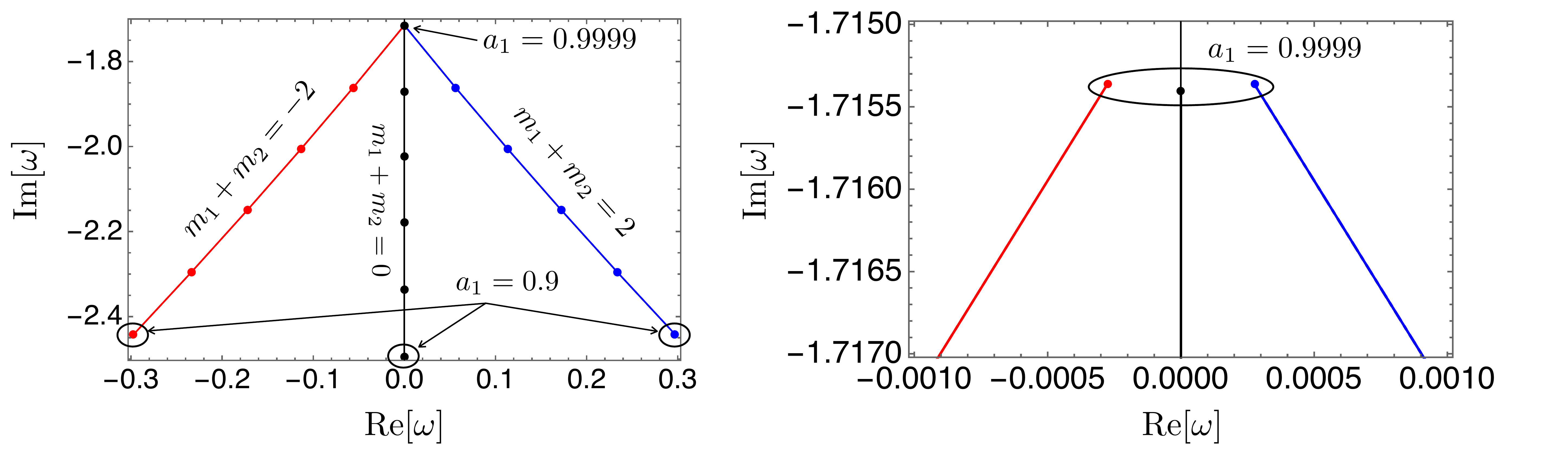}
 \caption{\footnotesize{The split of QN modes of $\ell=2$ for equal-rotation case ($a_1=a_2$) with $M=5, L=1, \mu=0.01$. 
 The range of the spin parameter is from $0.9$ to $0.9999$ with 
 the equal intervals: $0.01998$.
 The coloring is given as $m_1+m_2=-2$ (red), $m_1+m_2=0$ (black), $m_1+m_2+2$ (blue).
 The right panel 
 is an enlarged view of the left panel for the region around $a_1=a_2=0.9999$.
 }}
 \label{fig:QNM_a1=a2}
\end{figure}

\begin{figure}[H]
 \centering
 \includegraphics[width=0.9\linewidth]{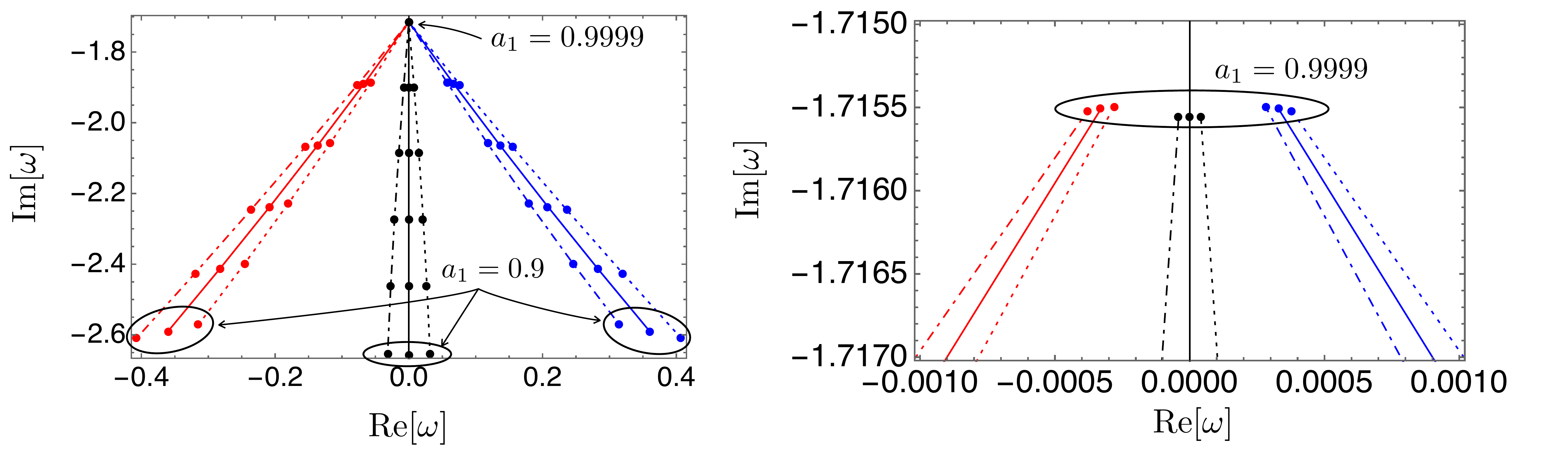}
 \caption{\footnotesize{QN frequencies 
 of $\ell=2$ modes for different-rotation case with fixed ratio ($1-a_2=1.4(1- a_1)$) with $M=5, L=1, \mu=0.01$. 
 We take $a_1$ as a variable and $a_2$ is given by the 
 above relation. The range of $a_1$ is from $0.9999$ to $0.9$ with the equal intervals: $0.01998$. 
 The coloring is the same as Fig.~\ref{fig:QNM_a1=a2}: 
 $m_1+m_2=-2$ (red), $m_1+m_2=0$ (black), $m_1+m_2+2$ (blue).
 The line type corresponds to the magnitude relation between $m_1$ and $m_2$: $m_1=m_2$ (solid), $m_1>m_2$ (dot-dashed), $m_1<m_2$ (dotted)}.
 }
 \label{fig:QNM_fixedratio}
\end{figure}

Let us focus on the longest-lived modes for each of the three sequences characterized by the value of $m_1+m_2$ in Fig.~\ref{fig:a09999_overtone}.
In Figs.~\ref{fig:QNM_a1=a2} and \ref{fig:QNM_fixedratio}, we show the splitting of these modes with large spin parameter for the equal-rotation case and 
different-rotation case, respectively. 
The equal-rotation case is depicted in Fig.~\ref{fig:QNM_a1=a2}.
Although the formulation with the local Heun function does not work
to compute the QN modes for the exactly extremal case, it seems that all the sequences converge to a common value.
On the other hand, Fig.~\ref{fig:QNM_fixedratio} shows the case of different rotations but the ratio between the two spin parameters is fixed 
as $1-a_2 = 1.4 (1-a_1)$.
There are nine sequences by reflecting the difference of the spin parameters. 
This splitting is the same behavior as the case of another sequence for the low spin case shown in Figs.~\ref{fig:QNMeq} and \ref{fig:QNMdif}.
While the absolute values of the imaginary parts of the QN modes in Figs.~\ref{fig:QNMeq} and \ref{fig:QNMdif} increase as 
the spin parameter increases, the imaginary parts for the QN modes in Figs.~\ref{fig:QNM_a1=a2} and \ref{fig:QNM_fixedratio} show the opposite dependency on the spin parameter.

\begin{figure}[H]
 \centering
 \includegraphics[width=0.85\linewidth]
 {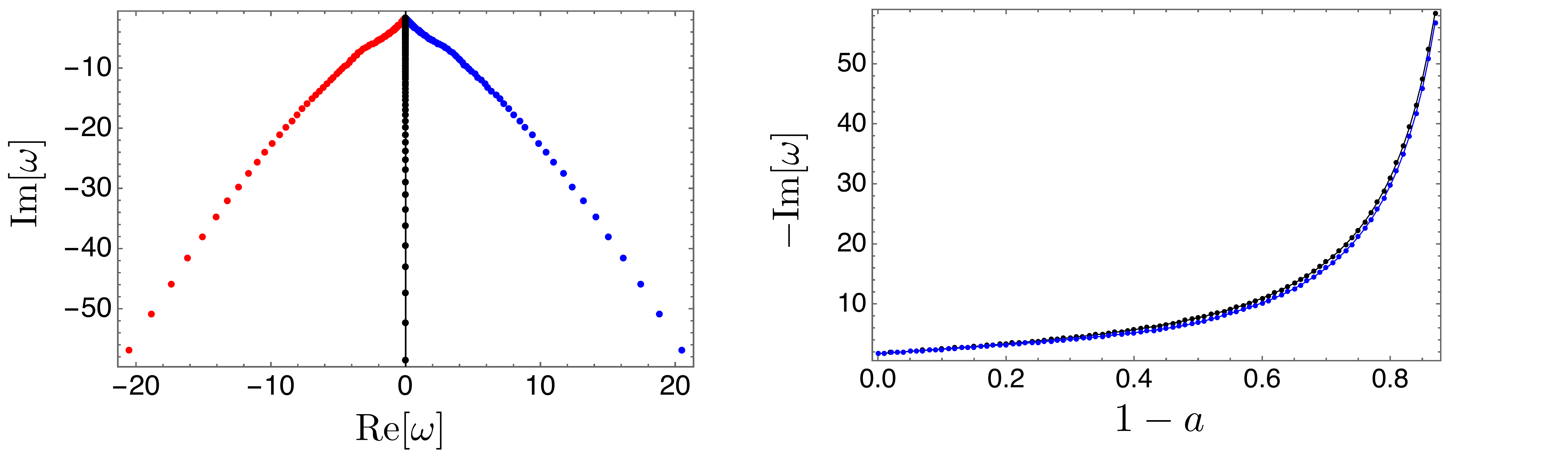}
 \caption{\footnotesize{
 QN frequencies of $\ell=2$ modes for equal-rotation case 
 $a_1=a_2=a$. The range of the spin parameter is 
 from $a=0.13$ to $a=0.9999$ with the interval $0.01$. 
 The coloring is given as 
 $m_1+m_2=-2$ (red), $m_1+m_2=0$ (black), $m_1+m_2+2$ (blue). 
 The right panel shows the spin parameter dependence 
 of the imaginary part of the purely 
 imaginary modes (black) and that of the modes on the both sides of the imaginary axis (blue), respectively.
}}
 \label{fig:QNMs}
\end{figure}

In Fig.~\ref{fig:QNMs}, we keep track of the longest-lived modes for each of the three sequences for the equal-rotation case shown in Fig.~\ref{fig:a09999_overtone}
with relatively wide range of the spin parameter: $a=0.13$, $0.14$, $\cdots$, $0.98$, $0.99$, $0.9999$. 
They are close to each other for the near extremal case, and move away as the spin parameter decreases. 
While their imaginary parts are almost the same for the near extremal case, as the spin parameter decreases, the purely imaginary mode 
has slightly larger amplitude as shown in the right panel of Fig.~\ref{fig:QNMs}.

It is interesting to note that a special behavior of the pure imaginary modes have been reported in different context. 
In \cite{Finazzo:2016psx}, the critical behavior of non-hydrodynamic QN modes is explored in the context of so-called 1-R charge black hole (1RCBH) model.
In Fig.~8 in \cite{Finazzo:2016psx}, it is shown that the purely imaginary mode goes to $-i\infty$ as $\mu/T$ reaches the threshold value $\mu/T=\pi/\sqrt{2}$, where $\mu$ and $T$ are the $U(1)$ R-charge chemical potential and temperature of the black brane background, respectively. 
The threshold $\mu/T=\pi/\sqrt{2}$ corresponds to a critical point of the phase diagram of the 1RCBH model.
A similar behavior is also studied in \cite{Atashi:2022ufl, Ishigaki:2021vyv} for the purely imaginary modes of the D7 brane black hole embedding solution.

It is nontrivial for the Kerr-AdS$_5$ case whether the purely imaginary modes may go to $-i\infty$ at some threshold spin parameter while the modes on the both sides of the imaginary axis remain finite.
If it is the case, the threshold spin would have some physical meaning.
However, we confirm that, at least down to $a=0.13$, the longest-lived purely imaginary mode exists with the finite imaginary part being almost the same as the modes on the both sides of the imaginary axis.
Nevertheless, the purely imaginary modes may or may not have the above characteristic behavior, if we consider more general case such as a Kerr-Newman-AdS$_5$ black hole. 
We hope to address this issue in future work.

\section{Greybody Factor and Superradiance}
\label{sec:greybody}

In this section, we consider a wave scattering problem in the Kerr-AdS$_5$ spacetime to define the greybody factor, which is a transmittance of the Hawking radiation from the black hole. Naively, similar to the case of asymptotically flat or de Sitter spacetime, one may think that it would be defined as the ratio of the coefficients in the UP mode, which would be given by:
\begin{align} 
\label{Rup-Heun-tort} 
R_{{\rm up}}(r) &= \begin{cases}
D^{\text{(up)}} \ e^{i\tilde{\omega}r_*}
+D^{\text{(ref)}}\ e^{-i\tilde{\omega}r_*}, & (r\to r_+) ,\\
D^{\text{(decay)}}\ r^{-\sigma}, & (r \gg r_+).
\end{cases}
\end{align} 
However, the asymptotic behavior of the Kerr-AdS$_5$ spacetime prevents us doing it in the sense that this boundary condition requires the decaying solution rather than the outgoing solution at the conformal infinity. 
Therefore, we take the other way to compute the greybody factor 
based on the idea adopted in 
\cite{Harmark:2007jy,Rocha:2009xy,Jorge:2014kra}.
To this end, we consider IN mode instead of the UP mode in the present spacetime by defining an 
ingoing and outgoing waves and evaluate the reflection coefficient. 
Then, subtracting the reflection coefficient from the unity, the greybody factor 
can be evaluated \cite{Harmark:2007jy,Rocha:2009xy,Jorge:2014kra}.
Strictly speaking, this flipping method between 
IN and UP mode exactly holds only for the asymptotically flat or de Sitter spacetime because the 
effective potential defined by the radial equation has similar behavior at the black hole horizon and at the conformal infinity or cosmological horizon, respectively. 
However, the method works for the evaluation of the greybody factor in the case that
it is possible to identify the ingoing and outgoing waves in a far region from the black hole horizon.

\subsection{Radial equation in the tortoise coordinate and the effective potential}
\begin{figure}[H]
 \centering
 \includegraphics[width=0.96\linewidth]{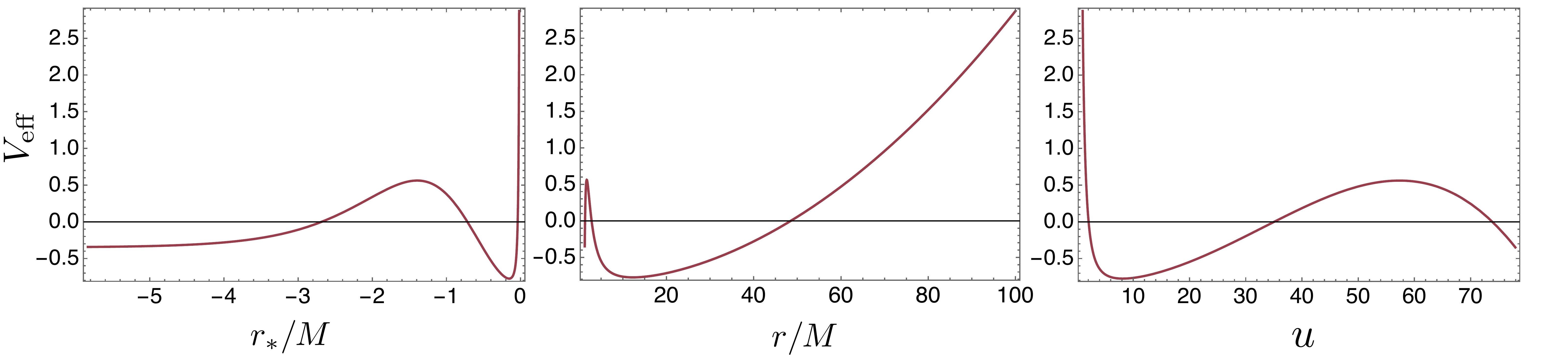}
 \caption{\footnotesize{The effective potential~$V_\text{eff}$ as a function of three different radial coordinates $r_*$, $r$, and $u=\omega L^2/r$ for the parameter set $L/M=10$, $a_1=0.5$, $a_2=0.25$, $\ell=2$, $m_1=1$, $m_2=1$.}}
 \label{fig:Veff}
\end{figure}

First, to clarify 
the wave behavior in the asymptotically AdS spacetime, let us rewrite the radial equation \eqref{KGrad} in the tortoise coordinate \eqref{tort} and introduce an effective potential for the scalar wave. 
Rescaling the radial function $R(r)$ in \eqref{KGrad} as 
\begin{equation}
\label{R_calY}
    R=H {\cal{Y}},\qquad  H\coloneqq \left[\dfrac{r}{(r^2+a_1^2)(r^2+a_2^2)}\right]^{1/2},
\end{equation}
and using the tortoise coordinate \eqref{tort}, Eq.~\eqref{KGrad} yields
\begin{align}
\label{sch}
    &\dfrac{d^2 {\cal{Y}}}{d r_*^2} - V_\text{eff}{\cal{Y}}=0, \\
   \notag & V_\text{eff}=-r^2H^4  \Delta^2\left[\dfrac{(r\Delta)'}{r\Delta}\df{H'}{H}+\left(\dfrac{H'}{H}\right)'+\left(\dfrac{H'}{H} \right)^2 \right]
    -\left[\omega-\dfrac{a_1m_1 \Xi_1}{r^2+a_1^2}-\dfrac{a_2 m_2 \Xi_2}{r^2+a_2^2}\right]^2\\
    &\quad \quad \quad \quad
    +\dfrac{r^2 \Delta}{(r^2+a_1^2)^2(r^2+a_2^2)^2}
    \left[r^2(\lambda +\mu^2 r^2) +(a_1 a_2 \omega -a_2 m_1 \Xi_1 -a_1 m_2 \Xi_2)^2\right].
\end{align}
In Fig.~\ref{fig:Veff}, we show the effective potential~$V_\text{eff}$ as a function of $r_*$, $r$, and a dimensionless 
inverse radial coordinate $u=\omega L^2/r$, 
which we shall make use of in the next subsection.
\noindent
As shown in Fig.~\ref{fig:Veff}, there exists a potential barrier as $r\to \infty$
which corresponds to the AdS boundary. 
Therefore, the wave scattering problem is quite different from that for 
asymptotically flat spacetimes.
Specifically, while the ingoing wave near the black hole horizon can be identified clearly since the potential is almost constant, 
it should be noted that the identification of the ingoing and outgoing waves 
around the local bottom near the AdS boundary is not always possible. 
As we will see in the following sections, it is legitimate for limited parameter regions.

\subsection{Greybody factor with far region approximation}
\label{ssec:Hankel}
As a first step, let us explain how to define the greybody factor based on the far region approximation with $r/L \gg 1$~\cite{Jorge:2014kra}.
With the far region approximation, we can transform the radial equation~\eqref{KGrad} into the Bessel's differential equation.
Introducing the dimensionless inverse radial coordinate $u=\omega L^2/r$, 
the radial equation~\eqref{KGrad} can be rewritten under the approximation $u \ll \omega L$ as
\be
u^2\df{d^2 R }{du^2}-3u\df{dR}{du}+\left[\left(1-\df{\lambda}{\omega^2 L^2}\right) u^2 -\mu^2 L^2\right]R=0
\ee
The linearly independent solutions of this differential equation are given in terms of the Bessel and Neumann functions, of which 
asymptotic behaviors are\footnote{To derive this form, we used the relation between the Bessel and Neumann functions:
\be
N_\nu (u)= \df{J_\nu(u)\cos{\nu \pi}-J_{-\nu}(u)}{\sin{\nu \pi}},
\ee
and their asymptotic form near $u=0$:
\be
J_{\nu}(a u)=\df{2^{-\nu}a^\nu}{\Gamma(1+\nu)}u^\nu,\qquad  
N_{\nu}(a u)=-\df{2^{-\nu} a^{\nu}\cos{(\pi \nu)}}{\pi}\Gamma(-\nu)u^\nu-\df{2^\nu a^{-\nu}}{\pi}\Gamma(\nu) u^{-\nu}.
\ee} 
\begin{align}
\label{Bessel}
R_\text{J}&=u^2J_{\sigma-2}(\sqrt{1-\lambda/(\omega^2L^2)}\ u)\simeq A_\text{J}\dfrac{1}{r^\sigma},\\
\label{Neumann}
R_\text{N}&=u^2N_{\sigma-2}(\sqrt{1-\lambda/(\omega^2L^2)}\ u)\simeq A_\text{N}\dfrac{1}{r^\sigma}+B_\text{N}r^{\sigma-4},
\end{align}
where 
\begin{align}
A_\text{J}&=\df{4\left(1-\frac{\lambda}{\omega^2 L^2}\right)^{\frac{\sigma-2}{2}}}{\Gamma(\sigma -1)} \left(\frac{\omega L^2}{2}\right)^\sigma,\\ 
A_\text{N}&=-\dfrac{\cos{(\pi \sigma)}\Gamma(2-\sigma)2^{2-\sigma}}{\pi}\left(1-\frac{\lambda}{\omega^2 L^2}\right)^{\frac{\sigma-2}{2}}(\omega L^2)^\sigma,\\ B_\text{N}&=-\dfrac{\Gamma(\sigma-2)2^{\sigma-2}}{\pi}\left(1-\frac{\lambda}{\omega^2 L^2}\right)^{\frac{2-\sigma}{2}}(\omega L^2)^{4-\sigma}.
\end{align}

To extract the in/outgoing properties of the solutions in the far region, it is more convenient to use another set of solutions based on the Hankel functions $H_{\sigma-2}^{(1)}=J_{\sigma-2}+iN_{\sigma-2}$ and $H_{\sigma-2}^{(2)}=J_{\sigma-2}-iN_{\sigma-2}$, since their asymptotic forms are nothing but the ingoing and outgoing waves, respectively.
Note that in order to define an appropriate wave scattering problem for introducing the notion of greybody factor with potential barrier around 
the AdS boundary, at least the following condition is necessary: 
The approximation is valid from the far region to the 
local bottom of $V_\text{eff}$, and the connection between the Hankel functions and 
the ingoing solution at the black hole horizon is possible. To satisfy the 
condition, the following additional restriction to the parameters is 
introduced in \cite{Jorge:2014kra}:
\be
\mu=0,\qquad \omega \ll T_+.
\label{cond_prev}
\ee

\begin{figure}[H]
 \centering
 \includegraphics[width=0.76\linewidth]{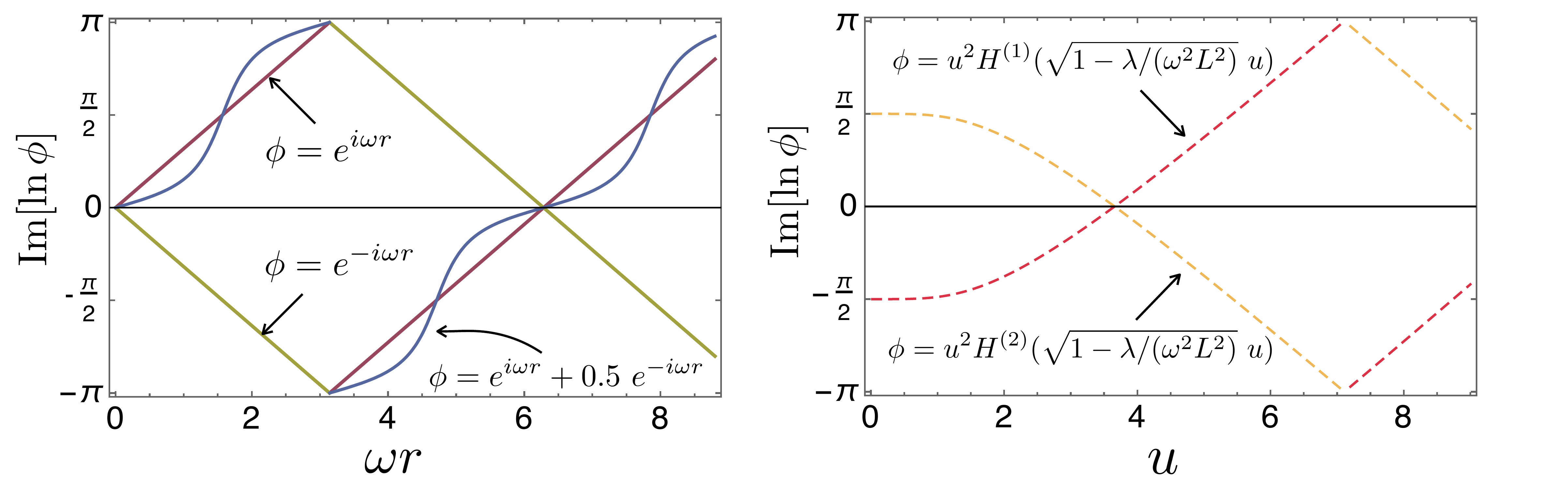}
 \caption{\footnotesize{The test function for plane waves for $r$ and the Hankel functions 
 for $u\propto 1/r$ with $M^{-1/2}L=10$, $M^{-1/2}a_1=0.5, M^{-1/2}a_2=0.25, \ell=2, m_1=1, m_2=1$. 
 Growing lines (curves) for $r$ represents outgoing wave, whereas decaying lines (curves) for $u$ ingoing waves. The yellow curves in the left panel 
 shows that the outgoingness of $e^{i\omega r}$ is partially broken by the term $0.5 e^{-i\omega r}$}.}
 \label{fig:inoutgoing_Hankel}
\end{figure}

To check the ``in/outgoingness'' explicitly, let us introduce a trial functional
\be
\text{Im}[\ln{\phi}],
\label{eq:Imlog}
\ee  
for a wave function $\phi$, which is the imaginary part of the phase of $\phi$.
To clarify the property of this trial functional, in Fig.~\ref{fig:inoutgoing_Hankel}, we show simple examples 
such as a superposition of plane waves (left panel) in addition to the case of the Hankel functions (right panel). 
By definition, if \eqref{eq:Imlog} grows (decays) linearly as a function of $r$, the input wave function~$\phi$ is purely outgoing (ingoing).
As shown in the left panel of Fig.~\ref{fig:inoutgoing_Hankel}, the deviation from the linear behavior indicates the breaking of its in/outgoingness. 
The right panel of Fig.~\ref{fig:inoutgoing_Hankel} depicts the independent solution of the wave equation in terms of the Hankel functions as a function of $u$.
At $u\to 0$, they approach constant since they are superposition of growing and decaying modes and hence the imaginary part of the phase remains constant.
On the other hand, the Hankel functions $H^{(1)}$ and $H^{(2)}$ indeed correspond to the ingoing and outgoing modes, respectively, 
around the local bottom of the effective potential shown in Fig.~\ref{fig:Veff}.
Note also that the behavior of the in/outgoing waves is opposite when we use the inverse radial coordinate $u$.
Namely, if \eqref{eq:Imlog} grows (decays) linearly as a function of $u$, the input wave function~$\phi$ is purely ingoing (outgoing). 

\noindent
Using the above basis of the solutions, the IN mode can be written as 
\begin{align} 
\label{Rin-Hankel} R_{{\rm in}}(r) &= \begin{cases}
e^{-i\tilde{\omega} r_*}, & (r\to r_+) ,\\
C_{1}u^2 H^{(1)}_{\sigma-2}(\sqrt{1-\lambda/(\omega^2 L^2) }\  u)+ C_{2}u^2 H^{(2)}_{\sigma-2}(\sqrt{1-\lambda/(\omega^2 L^2) }\  u), & (r \gg r_+),
\end{cases}
\end{align} 
and the greybody factor $\Gamma_{\ell m_1 m_2}(\omega)$ is defined by
\be
\label{greyJorge}
\Gamma_{\ell m_1 m_2}(\omega) =1-\left|\dfrac{C_2}{C_1} \right|^2.
\ee

To reiterate, the approach with the far region approximation works if the following condition is satisfied:
\begin{enumerate}
\renewcommand{\theenumi}{\alph{enumi}}
\renewcommand{\labelenumi}{{\bf (\theenumi):}}
\item \label{check-Hankel} The scalar field is massless, and the far region approximation with the Hankel functions works. Namely, the condition~\eqref{cond_prev} is satisfied.
\end{enumerate}

\subsection{Greybody factor with Heun function}
\label{ssec:Heun}
In this subsection, we construct a method to extract the ingoing and outgoing 
waves with the exact solution $R_{11}$ and $R_{12}$ in terms of the local Heun functions, 
and obtain the coefficients for the IN mode in more general regime.
Since we have exact solutions of the radial equation~\eqref{KGrad} 
around the regular singular points at 
$z=0\ (r\rightarrow r_+)$ and 
$z=1\ (r\rightarrow \infty)$, it is possible to 
write down the coefficients exactly, 
which we will denote as $C_\text{in}$ and $C_\text{out}$ here. 
They can be represented as linear combinations of the connection 
coefficients \eqref{C11}, as we shall see below.

We would like to construct the IN mode corresponding to \eqref{Rin-Hankel} with the exact solution in terms of the local Heun functions.
From the asymptotic behavior~\eqref{R01asym}, \eqref{R02asym}, it is clear that $R_{02}$ satisfies the purely ingoing condition at the black hole horizon.
From the connection formula~\eqref{y02rel}, $R_{02}$ is related to the local solutions around $z=1$ with the connection coefficients $C_{11}$ and $C_{12}$ given in \eqref{C11}.
Therefore, the IN mode is exactly written as 
\begin{align} 
\label{Rin-R11-R12} R_{{\rm in}}(r) &= \begin{cases}
R_{02}(r), & (r\to r_+) ,\\
C_{21}R_{11}(r)+ C_{22}R_{12}(r), & (r \gg r_+).
\end{cases}
\end{align}
As mentioned above, from the asymptotic behavior~\eqref{R11asym}, \eqref{R12asym}, we see that $R_{11}$ and $R_{12}$ are decaying and growing modes, respectively.
The situation here is similar to the one with the far region approximation, where we change the basis of the solutions from the Bessel and Neumann functions
to the Hankel functions, which describe in/outgoing modes.
Applying the same strategy, we can write down 
\begin{align} 
\label{Rin-Rp-Rm} R_{{\rm in}}(r) &= \begin{cases}
R_{02}(r), & (r\to r_+) ,\\
C_\text{out}R_{+}(r)+C_\text{in} R_{-}(r), & (r \gg r_+),
\end{cases}
\end{align}
with
\begin{align}
    R_-(r)=\dfrac{A_\text{J}+iA_\text{N}}{A_{11}}R_{11}(r)+i\dfrac{B_\text{N}}{A_{12}}R_{12}(r), \\
    R_+(r)=\dfrac{A_\text{J}-iA_\text{N}}{A_{11}}R_{11}(r)-i\dfrac{B_\text{N}}{A_{12}}R_{12}(r).
\end{align}
Note that $R_-$ and $R_+$ correspond to the Hankel functions $H^{(1)}$ and $H^{(2)}$, respectively. 
For defining these functions, we compared the asymptotic forms of the Bessel and Neumann functions [\eqref{Bessel} and \eqref{Neumann}] and those of $R_{11}$ and $R_{12}$ [\eqref{R11asym} and \eqref{R12asym}].
Comparing \eqref{Rin-R11-R12} and \eqref{Rin-Rp-Rm}, we obtain

\begin{align}
C_\text{in}&=\dfrac{A_{11}A_{12}}{2iA_\text{J} B_\text{N}}\left(i\dfrac{B_\text{N}}{A_{12}}C_{21}+\dfrac{A_\text{J}-iA_\text{N}}{A_{11}}C_{22}\right),\\
C_\text{out}&=\dfrac{A_{11}A_{12}}{2iA_\text{J} B_\text{N}}\left(i\dfrac{B_\text{N}}{A_{12}}C_{21}-\dfrac{A_\text{J}+iA_\text{N}}{A_{11}}C_{22}\right).
\end{align}
As the radial equation \eqref{KGrad} is the Strum-Liouville type equation, the following quantity with the Wronskian is conserved:
\be
\label{wronskian}
r\D W_r[R_1,R_2]=\text{const}, \quad \quad 
\ee
where $W_r[R_1,R_2]=R_1 \frac{dR_2}{dr}-R_2\frac{dR_1}{dr}$. Using this, we obtain the greybody factor corresponding to \eqref{greyJorge} in terms of the connection coefficients for the local Heun functions 
as
\be
\label{grey-Heun}
\Gamma_{\ell m_1 m_2}(\omega) =1-\left|\dfrac{C_\text{out}}{C_\text{in}}\right|^2=1-\dfrac{\left|i\dfrac{B_\text{N}}{A_{12}}C_{21}-\dfrac{A_\text{J}+iA_\text{N}}{A_{11}}C_{22} \right|^2}{\left|i\dfrac{B_\text{N}}{A_{12}}C_{21}+\dfrac{A_\text{J}-iA_\text{N}}{A_{11}}C_{22} \right|^2}.
\ee
As we have seen, $A_{11}$, $A_{12}$, $A_\text{J}$, $A_\text{N}$, $B_\text{N}$ are written down explicitly, whereas $C_{21}$, $C_{22}$ are written as the ratio of the Wronskians between local Heun functions.

\begin{figure}[H]
 \centering
 \includegraphics[width=0.5\linewidth]{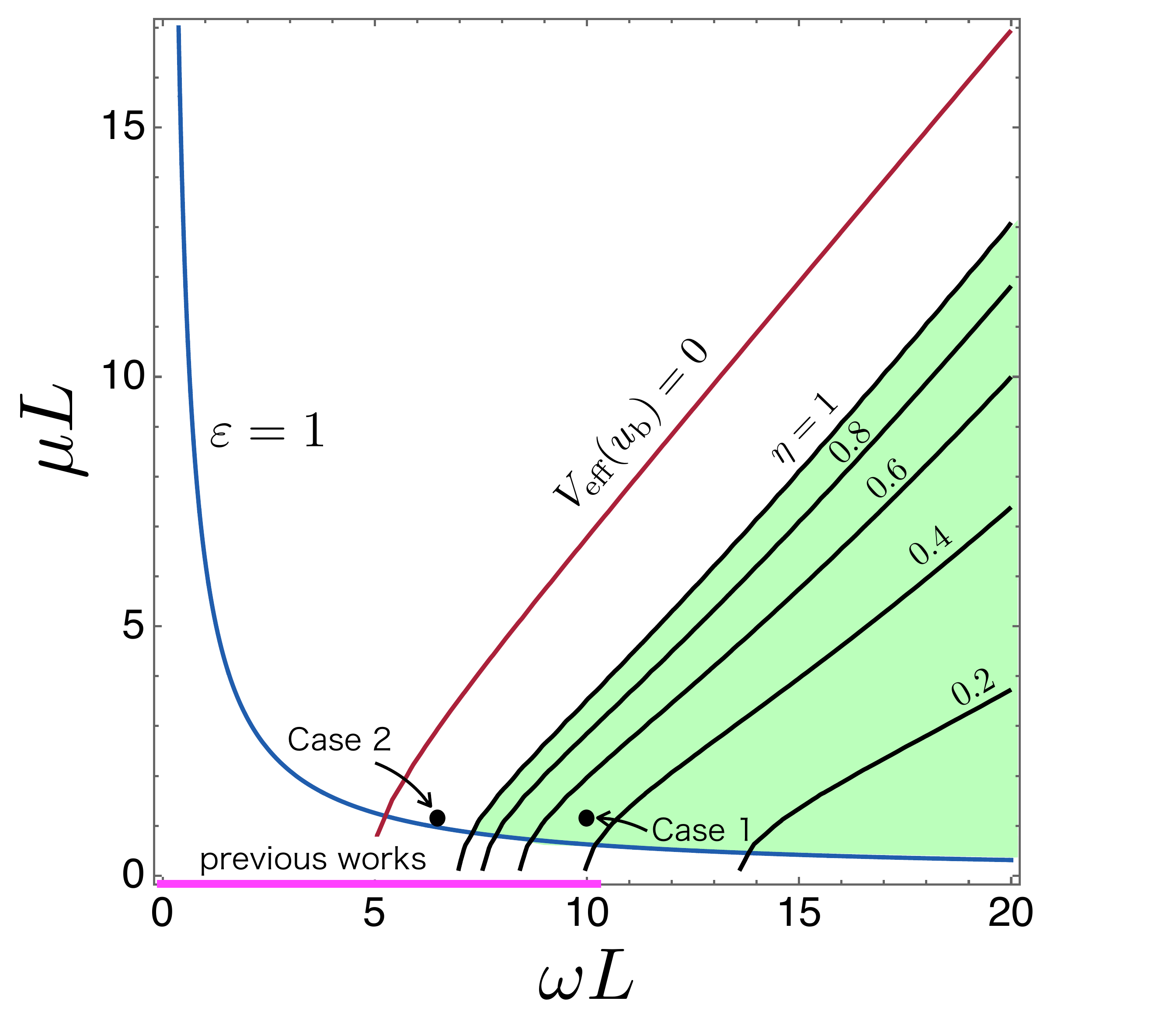}
 \caption{\footnotesize{Contours of $\varepsilon$, 
 $\eta$, and $V_\text{eff}(u_\text{b})$ for 
 $M=1, M^{-1/2}a_1=0.5, M^{-1/2}a_2=0.25, \ell=2, m_1=1, m_2=1$. 
 Our method can be utilized for parameter sets in the green shaded region. The magenta thick line on $\omega L$ axis depicts the parameter region~\eqref{cond_prev} where the method based on the far region approximation works. Its cutoff is given by $T_+ L=10.2$. 
 Parameter sets corresponding to Case 1 and Case 2 are marked with black dots.}}
 \label{fig:region}
\end{figure}

For our method to work well, it is necessary that $R_-$ and $R_+$ behave as ingoing and outgoing waves, respectively.
Combined with the condition on the mass of the scalar field explained at the end of \S\ref{ssec:rad}, 
the condition under which our method works can be stated as follows:
\begin{enumerate}
\renewcommand{\theenumi}{\alph{enumi}}
\renewcommand{\labelenumi}{{\bf (\theenumi):}}
\setcounter{enumi}{1}
\item \label{check-Heun} The scalar field is massive $\mu\ne 0$, and out/ingoingness of the exact solution $R_\pm$ holds at some finite range around the local bottom of the effective potential near the AdS boundary.
\end{enumerate}
Ultimately, out/ingoingness of $R_\pm$ should be checked by, e.g., the imaginary part of the phase~\eqref{eq:Imlog}.
However, it would be also useful to have analytical criteria.
Rough criteria for our method to work can be written as
\be \label{criteria}
\varepsilon \coloneqq \df{2\pi }{ \omega \mu L^2} \ll 1,\qquad 
\eta \coloneqq \df{2\pi}{\omega} \left|\df{V_\text{eff}^{''}(u_\text{b})}{V_\text{eff}(u_\text{b})} \right|^{\frac{1}{2}} \ll 1, \qquad 
V_\text{eff}(u_\text{b}) < 0,
\ee
where 
$u_\text{b}$ is the position of the local bottom of 
$V_\text{eff}$ and $''$ represents the second derivative with respect to $r$. 
Physical meaning of each inequality is as follows.
The first inequality ensures that the wavelength and the 
Compton length of the scalar field are sufficiently smaller than the AdS scale, whereas the second inequality stems from the condition that the curvature of the local bottom of the effective potential 
is much smaller than the wavelength.
In other words, the effective potential is sufficiently flat around the local bottom. 
The third condition, the negativity of $V_\text{eff}(u_\text{b})$ 
ensures the existence of propagating wave modes around the local bottom.
In Fig.~\ref{fig:region} we show the parameter region where the criteria~\eqref{criteria} are satisfied.
We note that, at least for the parameter region of our interest, the condition~$V_\text{eff}(u_\text{b}) < 0$ is always satisfied if $\eta\ll 1$ is satisfied.
Our method works well for relatively large $\omega L$ and small $\mu L$ but not for the exactly massless case $\mu L=0$, for which the method with the far region approximation works.
\noindent
Once more, note that the evaluation of the allowed parameter region is just rough criteria.
To conclude whether our method works or not for a given parameter set and to define a proper wave scattering problem, it is necessary to check 
whether the linear behavior of the trial functional~\eqref{eq:Imlog} of $R_\pm$ holds at some finite range around the local bottom of the effective potential.  

Now, let us compare the two approaches: the one with the far region approximation described in \S\ref{ssec:Hankel}, and the one with the Heun function.
The far region approximation works for the low-frequency massless scalar wave, and the upper bound~\eqref{cond_prev} on the frequency is determined by the Hawking temperature.
On the other hand, it is worthwhile to note that our method does not have an upper bound for the range of validity in the frequency domain, as shown in Fig.~\ref{fig:region}.
Actually, for larger $\omega$, our method works better. 
Therefore, once one confirms that the condition \eqref{check-Heun} is satisfied for small $\omega$, one can apply our method for larger $\omega$ too.

It is clear that the two conditions~\eqref{check-Hankel} and \eqref{check-Heun} are exclusive.
As shown in Fig.~\ref{fig:region}, the parameter regions of the validity of each of the two methods do not overlap.
In Fig.~\ref{fig:inoutgoing_Hankel}, we demonstrated the case where the far region approximation works, 
but for the massless case the solution~\eqref{y12def} with the local Heun function does not exist as we explained at the end of \S\ref{ssec:rad}.
Therefore, in this case, it is not possible to compare the two methods quantitatively.

In the rest of this subsection, we compare the two methods for the remaining two cases:
\begin{description}
    \item[Case 1] The method with the Heun function works, but the method with the Hankel fuctions does not [\eqref{check-Heun} is satisfied, but not \eqref{check-Hankel}].
    \item[Case 2] Neither method works [\eqref{check-Hankel} and \eqref{check-Heun} are not satisfied]. 
\end{description}
For each case, we pick up a representative parameter set, which is given below, and is shown in Fig.~\ref{fig:region}.
We shall plot the effective potential as a function of the inverse radial coordinate $u$ to clarify the position of the local bottom.
Then, around the local bottom, we investigate the trial functional \eqref{eq:Imlog} to check if the solutions can be identified as in/outgoing waves. 
We also plot $R_{\pm}$ composed of the local Heun functions and compare them to the Hankel functions.

Hereafter, we set $M=1$ and all other parameters with dimension are normalized by $M$.

\subsubsection*{Case 1}
\begin{figure}[H]
 \centering
 \includegraphics[width=0.7\linewidth]{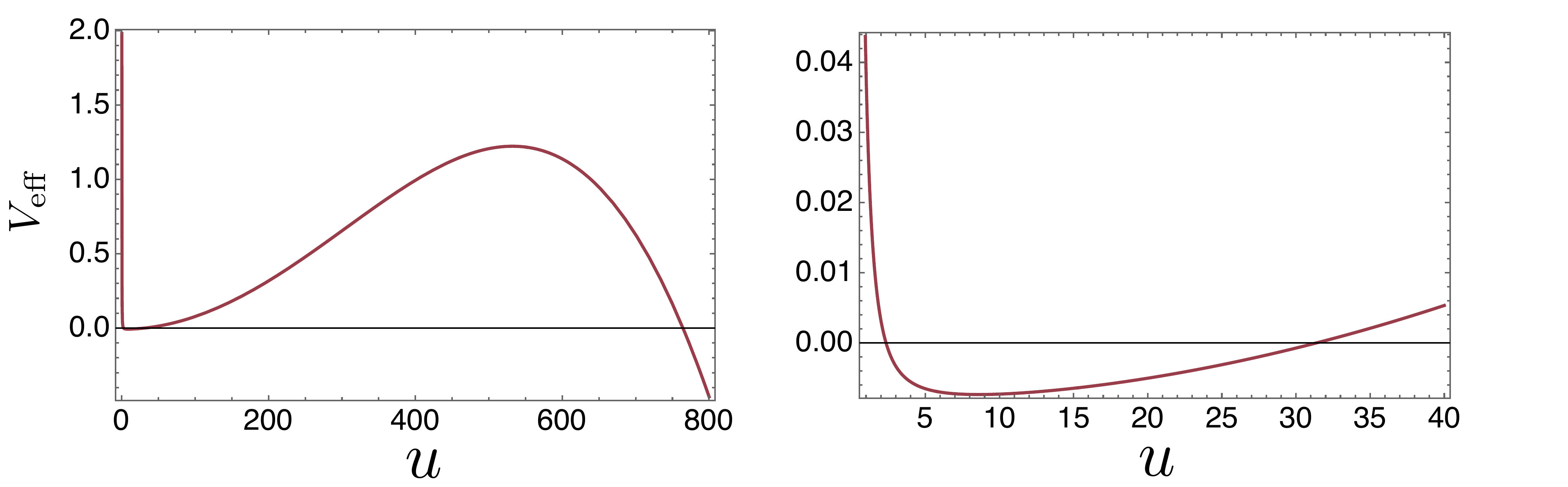}
 \caption{\footnotesize{The effective potential as a funciton of $u$ for the parameter set $L=100, a_1=0.5, a_2=0.25, \ell=2, m_1=1, m_2=1,\omega=0.1,\mu=0.01$. The right panel is the close-up plot around the local bottom. 
 Note that small $u$ corresponds to large $r$, and the 
 AdS boundary is located at $u=0$.}}
 \label{fig:Veff_HankelNG_oursOK}
\end{figure}

\begin{figure}[H]
 \centering
 \includegraphics[width=0.98\linewidth]{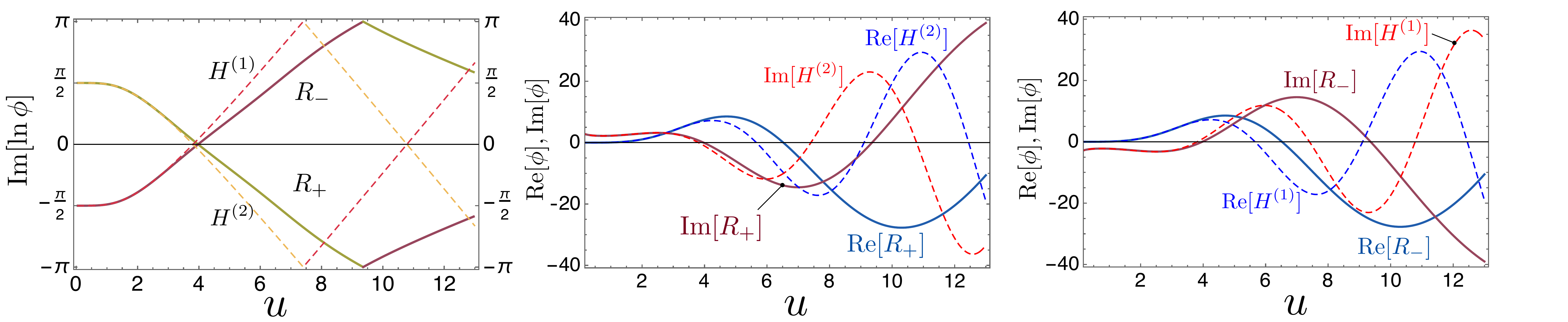}
 \caption{\footnotesize{The trial functional (left) and the Hankel functions and exact solutions $R_\pm$ (middle and right) 
 for $\ell=2,m_1=1,m_2=1$ mode. 
 The left panel shows the linear behavior of $R_\pm$ holds around the local bottom. In the middle and right panels, one can see 
 the difference between the Hankel functions and the exact solutions cannot be ignored for large $u$ around the local bottom.}}
 \label{fig:HankelNG_oursOK}
\end{figure}

For the Case 1, we choose a representative set of the parameters as $L=100$, $a_1=0.5$, $a_2=0.25$, $\ell=2$, $m_1=1$, $m_2=1$, $\omega=0.1$, $\mu=0.01$. 
In Fig.~\ref{fig:Veff_HankelNG_oursOK}, we show the effective potential as a function of $u$.
We see that there is a wide plateau with negative potential value, and expect that the in/outgoing waves are solutions in this region.
For these parameters, $\varepsilon=0.63$, $\eta=0.45$.
Hence, the criteria~\eqref{criteria} are satisfied, which is also a good signal.
Indeed, we see that
the linear behavior of the imaginary part of the phase of $R_\pm$ holds around the local bottom from the first panel of Fig.~\ref{fig:HankelNG_oursOK}.
Although the tilt of the imaginary part of the phase is not constant, reflecting the slow variation of the effective potential around the local bottom, we can observe that $R_\pm$ exhibit periodic oscillations.
Hence, the condition~\eqref{check-Heun} is satisfied.
However, \eqref{check-Hankel} is not satisfied since the deviation of the Hankel 
functions from the exact solutions cannot be ignored and are no longer valid as approximated solutions.
As shown in the first panel in Fig.~\ref{fig:HankelNG_oursOK}, the phases of the Hankel function are deviated from the ones for the exact solutions.
The functions themselves are also deviated from the exact solutions, as shown in the second and third panels in Fig.~\ref{fig:HankelNG_oursOK}.

\begin{figure}[H]
 \centering
 \includegraphics[width=0.95\linewidth]{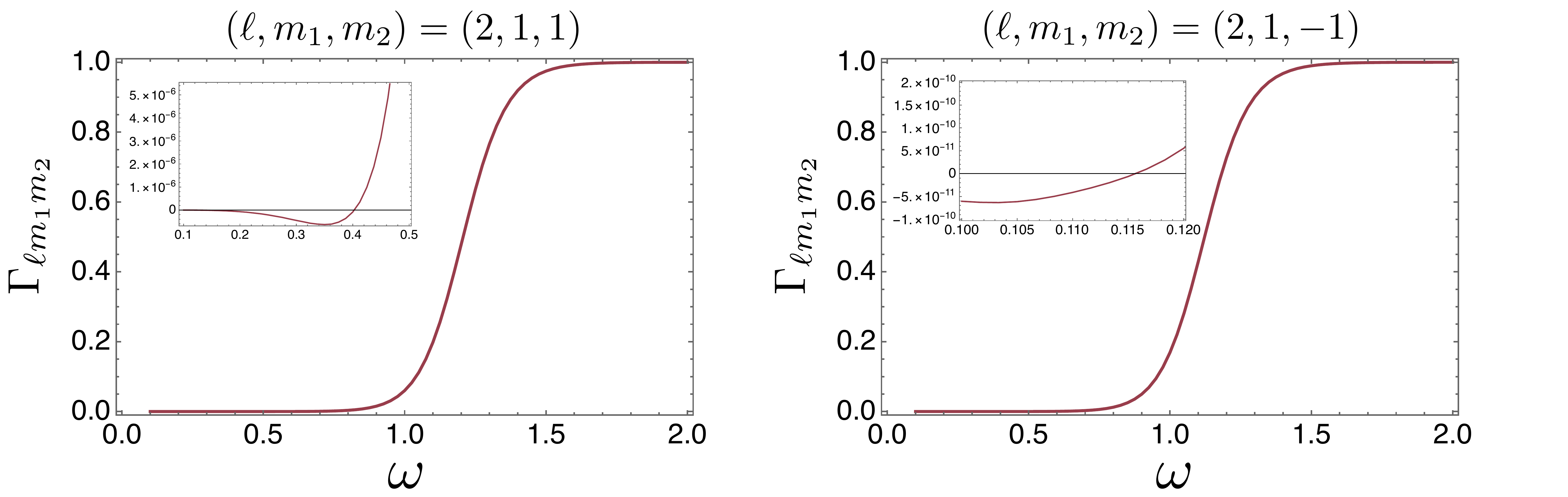}
 \caption{\footnotesize{Greybody factors with superradiance;  For the present spin parameters $a_1=0.5$ and $a_2=0.25$, the critical frequencies for superradiance given by $m_1 \Omega_{+,1}+m_2 \Omega_{+,2}$ are $0.4$ for $(\ell,m_1,m_2)=(2,1,1)$ and $0.116$ for $(\ell,m_1,m_2)=(2,1,-1)$, respectively. 
For frequencies near the transition of 
 the greybody factor from 0 to 1, the peak of the effective potential becomes almost zero.}}
 \label{fig:GB_superrad}
\end{figure}

As mentioned above, since we confirm that the method works well for $\omega=0.1$, we can also apply the method for high frequency regime.
In Fig.~\ref{fig:GB_superrad}, we show the greybody factor for $(\ell,m_1,m_2)=(2,1,1)$ and $(2,1,-1)$ modes for $0.1\leq \omega \leq 2$.
As expected, it transitions from $0$ to $1$ as the frequency increases.
We can also see superradiance in the low frequency region as the negative value of the greybody factor.
For the present spin parameters, the critical frequencies for superradiance given by $m_1 \Omega_{+,1}+m_2 \Omega_{+,2}$ are $0.4$ for $(\ell,m_1,m_2)=(2,1,1)$ and $0.116$ for $(\ell,m_1,m_2)=(2,1,-1)$, respectively, which are consistent with Fig.~\ref{fig:GB_superrad}. 
Note that we have checked out/ingoingness for $(\ell,m_1,m_2)=(2,1,-1)$ case as well before computing the greybody factor although we show $(\ell,m_1,m_2)=(2,1,1)$ in Fig.~\ref{fig:HankelNG_oursOK}. 

On the other hand, the method with the far region approximation only works at the low frequency regime.
For the present case, the allowed region is $\omega < T_+= 0.102$.
This is not sufficient to cover the frequency region where the greybody factor transitions from $0$ to $1$.

\subsubsection*{Case 2}
\begin{figure}[H]
 \centering
 \includegraphics[width=0.7\linewidth]{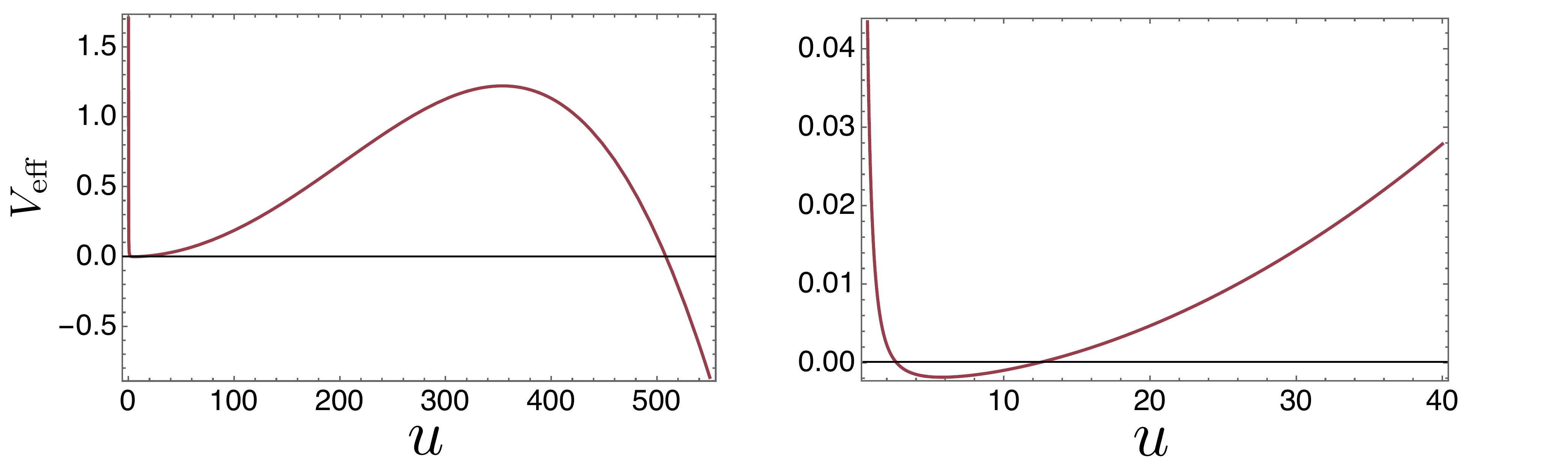}
 \caption{\footnotesize{The effective potential as a 
 function of $u$. The right panel is the close-up plot around the local bottom. The flat 
 region around the local bottom is slightly narrower than $V_\text{eff}$ in the Case 1 (Fig.~\ref{fig:Veff_HankelNG_oursOK})}}
 \label{fig:Veff_HankelNG_oursNG}
\end{figure}

\begin{figure}[H]
 \centering
 \includegraphics[width=0.98\linewidth]{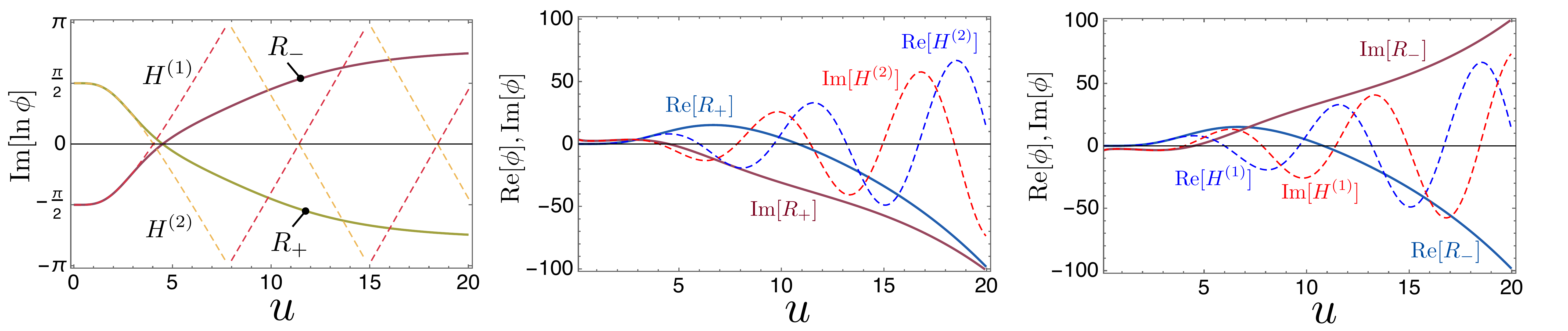}
 \caption{\footnotesize{The trial functional and explicit plots of the Hankel functions and exact solutions $R_\pm$. The first panel shows the linear behavior of $R_\pm$ breaks down around the local bottom. Moreover, the difference between the Hankel functions and the exact solutions is quite large. Therefore, 
 neither method works.}}
 \label{fig:HankelNG_oursNG}
\end{figure}

For the Case 2, we choose a representative set of the parameters as $L=100$, $a_1=0.5$, $a_2=0.25$, $\ell=2$, $m_1=1$, $m_2=1$, $\omega=0.067$, $\mu=0.01$. 
For these parameters, $\varepsilon=0.93$, $\eta=1.34$, and the flat region around the local bottom of the effective potential in Fig.~\ref{fig:Veff_HankelNG_oursNG} is narrower than that of Fig.~\ref{fig:Veff_HankelNG_oursOK}, so we expect that $R_\pm$ do not behave as out/ingoing waves.
Indeed, in this case, \eqref{check-Heun} is not satisfied since the linear behavior of $R_\pm$ breaks down around the local bottom as shown in the first panel of Fig.~\ref{fig:HankelNG_oursNG}.
Also, the second and third panels of Fig.~\ref{fig:HankelNG_oursNG} indicate that the errors of 
the Hankel functions are quite large for larger $u$.

\section{Hawking Evaporation of Kerr-AdS$_5$}
\label{sec:evap}
In this section, we discuss the evaporation of a Kerr-AdS$_5$ black hole via Hawking radiation with massive scalar field for different-rotation case. In general, Hawking radiation includes several spin fields other than scalar field: graviton, photon, and neutrino(s) \cite{Page:1976df,Page:1976ki,Page:1977um}. 
Therefore, to obtain the evaporation rate, it is necessary to take into account the Hawking radiation of these fields.
Here, as a first step toward the rigorous evaluation of the evaporation rate, we focus on the contribution of the scalar field to the evaporation.
If the master equation of the other fields also take the form of the Heun's equation, it is straightforward to extend the following argument to include those fields.
Another caveat is that the scalar field we consider is not massless due to the choice of the set of local Heun function as mentioned in \S\ref{ssec:rad}. 
The mass introduces a lower cutoff on the energy that an emitted particle may have, 
which eliminates the lower-energy part of the spectrum that a massless particle would have~\cite{Page:1977um}. 
Therefore, nonzero rest masses of the emitted particles can have a large impact upon the evaporation rates unless $\mu/T_+\ll 1$. 
As we shall see below, in the present case, the lower cutoff is sufficiently small, and we can evaluate the contribution from the scalar field.

\begin{figure}[H]
 \centering
 \includegraphics[width=0.5\linewidth]{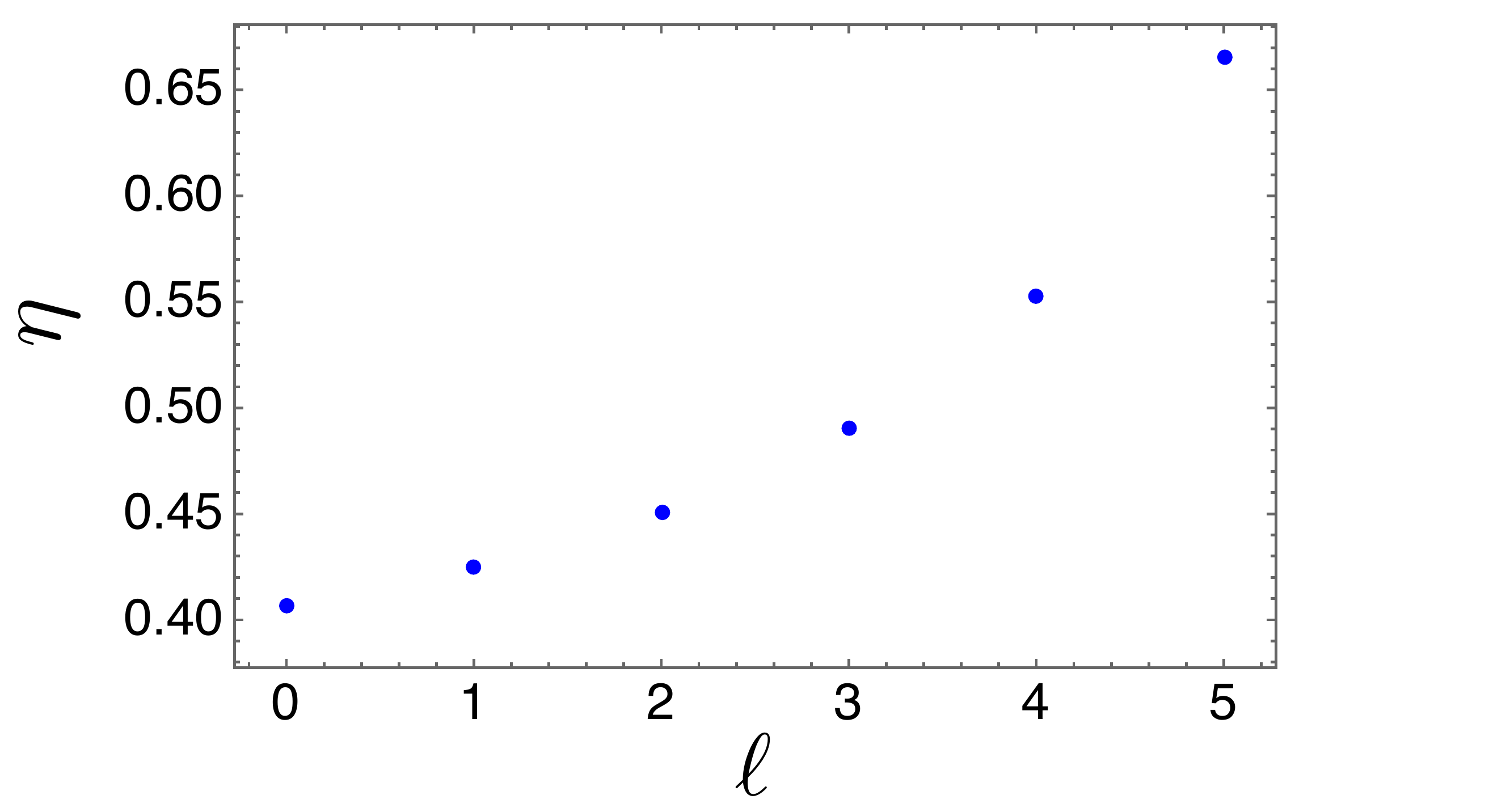}
 \caption{\footnotesize{$\ell$-dependence of $\eta$ for $a_1=0.5$, $a_2=0.25$, $\omega=0.1$, $\mu=0.01$, $m_1+m_2=\ell$ case. $\eta$ becomes gradually large for 
 $\ell$.}}
 \label{fig:eta_ell}
\end{figure}

For asymptotically AdS black holes, the number of emitted particles as the Hawking radiation is given by plugging the Hawking temperature into 
$1/(e^{\tilde{\omega}/T_+}-1)$~\cite{Hemming:2000as,Saraswat:2020zzf}, which is the same as the asymptotically flat case 
\cite{Page:1976df,Page:1976ki,Page:1977um}.
Using this particle number spectrum and the greybody 
factor, we define the spectrum of mass evaporation and angular momentum evaporation via Hawking radiation as 
\be
\label{evaporationrates}
    \dfrac{d^2 {\cal{M}}}{dt d\omega}
    =\dfrac{-1}{2\pi}\sum_{\ell m_1 m_2}\dfrac{\omega \Gamma_{\ell m_1 m_2}(\omega)}{e^{\tilde{\omega}/T_+}-1}, \quad
    \dfrac{d^2 {\cal{J}}_1}{dt d\omega}
    =\dfrac{-1}{2\pi}\sum_{\ell m_1 m_2}\dfrac{m_1 \Gamma_{\ell m_1 m_2}(\omega)}{e^{\tilde{\omega}/T_+}-1}
    ,\quad
    \dfrac{d^2 {\cal{J}}_2}{dt d\omega}
    =\dfrac{-1}{2\pi}\sum_{\ell m_1 m_2}\dfrac{m_2 \Gamma_{\ell m_1 m_2}(\omega)}{e^{\tilde{\omega}/T_+}-1},
\ee
Basically, our method becomes more invalid for lower frequency. 
Therefore, we first check the lower bound of the allowed parameter region using the scheme introduced in the previous subsection for small $\omega$. 
Also a systematic way to check if our method works for all possible angular and magnetic quantum numbers $(\ell,m_1,m_2)$ is necessary. 
Fig.~\ref{fig:eta_ell} depicts $\ell$-dependence 
of $\eta$ for a fixed $\omega$. As $m_1, m_2$-dependence 
is very weak, we plot $m_1+m_2=\ell$ case here. 
From Fig.~\ref{fig:eta_ell}, we see that $\eta$ becomes larger for higher $\ell$ modes. If we need to consider $\ell$ 
up to $\ell_\text{max}$\footnote{The cutoff $\ell_\text{max}$ is 
determined by confirming the contribution of $\ell_\text{max}$ 
modes to an evaporation rate is small enough to be ignored compared to that of lower $\ell$ modes.}, 
at first, $\eta$ for $\ell_\text{max}$ with small $\omega$ should be checked to search the lower bound 
frequency $\omega_{\rm min}$ for which $\eta$ remains to be so small
that $R_{\pm}$ are out/ingoing waves. 
Once we find the lower bound $\omega_{\rm min}$ for 
$\ell_\text{max}$, $\eta$ is sufficiently small for all other 
$(\ell,m_1,m_2)$ up to $\ell_\text{max}$.
Then, we can compute greybody factors up to 
$\ell=\ell_\text{max}$ correctly.

Here, we choose the parameters in the Case 1 in the previous 
section and consider $\omega$ in the range 
$0.1\leq \omega \leq 4.0$. For $\omega$ that is allowed 
in our method, we can compute the spectra of mass and spin evaporations, and 
obtain plots as functions of $\omega$. 
If those spectra converge to zero at the endpoints of the allowed region in the frequency domain, we can numerically integrate the spectrum to obtain evaporation rates $d{\cal{M}}/dt$, 
$d{{\cal{J}}_1}/dt$, and $d{{\cal{J}}_2}/dt$. 
As we mentioned above, the nonzero mass introduces a lower cutoff to the integral in the frequency domain.
However, the mass parameter here is $\mu=0.01$ and it is smaller than the Hawking temperature as $\mu/T_+\approx 0.1$, so the effect of the rest mass is expected to be not so significant.
Moreover, $\mu=0.01$ is smaller than the lower bound of the range of the allowed frequency in our method. 
Therefore, we integrate spectra in the range $0.1 \leq \omega \leq 4.0$.

\begin{figure}[H]
 \centering
 \includegraphics[width=0.86\linewidth]{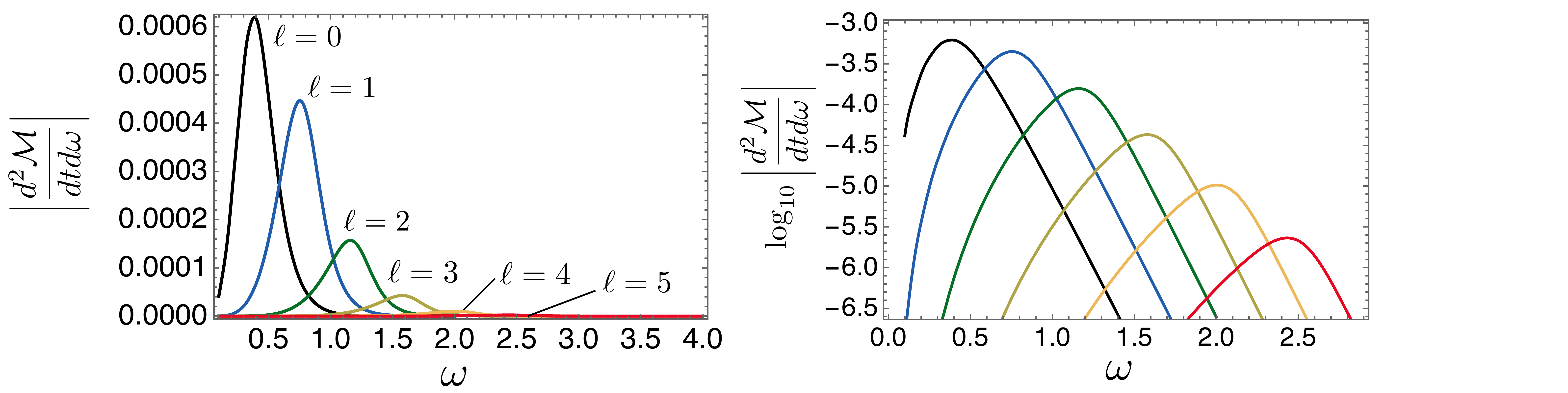}
 \caption{\footnotesize{Spectra of mass evaporation for 
 $\ell=0,1,2,3,4,5$ with $L=100$, $a_1 =0.5$, $a_2 =0.25$, $\mu =0.01$. The right panel depicts $\log_{10}{\left|\frac{d^2 {\cal{M}}}{dt d\omega}\right|}$ around the peak.}}
 \label{fig:mass_rate}
\end{figure}
\begin{figure}[H]
 \centering
 \includegraphics[width=0.9\linewidth]{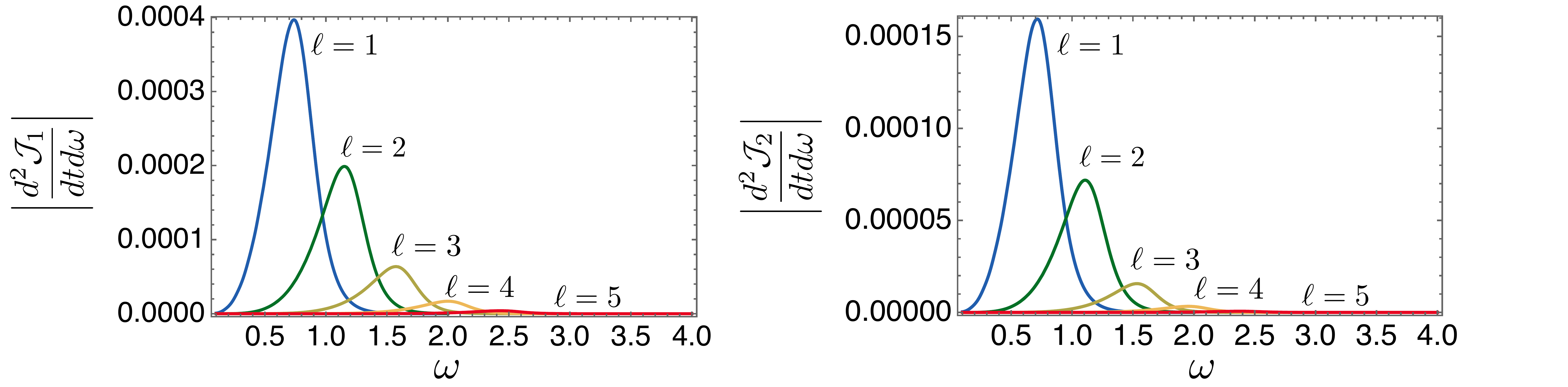}
 \caption{\footnotesize{Spectra of angular momentum evaporation for $\ell=1,2,3,4,5$ with $L=100$, $a_1=0.5$, $a_2 =0.25$, $\mu=0.01$.}}
 \label{fig:angular_rate}
\end{figure}
\noindent 
As shown in Figs.~\ref{fig:mass_rate} and \ref{fig:angular_rate}, in the spectra of Hawking evaporation \eqref{evaporationrates}, 
the contributions of the lower $\ell$ modes 
are relatively large and those of $\ell=5$ modes are
very small.
After integration in the frequency domain, the contributions of $\ell=5$ modes are about $0.5$\% of the summation up to $\ell=5$. 
Therefore, for the present case with $a_1=0.5$, $a_2=0.25$, 
adding up to $\ell=5$ modes is enough 
to compute the evaporation rates. 
Even around the most contributed frequency to 
the spectrum of mass evaporation, $\varepsilon$ and $\eta$ are sufficiently small for our method to work, that is,
the out/ingoingness of $R_\pm$ holds. 
Specifically, $(\varepsilon,\eta)=(0.015,0.117)$ for $\ell=0$, $\omega=0.4$, $(\varepsilon,\eta)=(0.007,0.028)$ for $\ell=1$, $\omega=0.8$, and $(\varepsilon,\eta)=(0.005,0.011)$ for $\ell=2$, $\omega=1.25$. This tendency holds for 
the spectra of angular momenta evaporation. Then, 
integrating over $\omega$, the values of 
evaporation rates are obtained as
\be \f{d{\cal{M}}}{dt}=-5.2\times 10^{-4},\quad 
\f{d{\cal{J}}_1}{dt}=-3.1\times 10^{-4},\quad
\f{d{\cal{J}}_2}{dt}=-1.1\times 10^{-4}. \ee
This result indicates that the larger angular momentum evaporates faster. 

Again, although the lower bound of those integrals should be given by the mass, $\omega=\mu=0.01$, we take the lower bounds as $\omega=0.1$ since 
the allowed frequency for the present parameter set is $0.1 \leq \omega \leq 4.0$. 
Around $\omega=0.1$, the spectra of the angular momenta is sufficiently close to zero, but the mass spectrum is not due to the contribution from the $\ell=0$ mode as shown in Fig.~\ref{fig:mass_rate}. 
Therefore, the low-frequency contribution to the mass evaporation is not taken into account precisely, while the contribution from the low-frequency tail is expected to be not so large.

Before closing this section, let us make comments on 
the mass of the scalar field and black hole. 
Here, as a demonstration, we choose $\mu=0.01$ with $L=100$, but 
it is possible to take much smaller mass parameter 
as long as $\mu L$ is small enough, which admits out/ingoingness of $R_\pm$. Moreover, although 
we set $M=1$ and consider a small black hole here, 
our method can also be applied to large black hole case 
as long as the existence of in/outgoing waves at 
a far region is guaranteed. In that case, 
a modulation due to the resonance reflecting the AdS scale would be expected.

\section{Conclusion}
\label{sec:con}

In this paper, we investigated the QN frequencies, greybody factor, and evaporation rates for the test massive scalar field in the Kerr-AdS$_5$ spacetime by employing the exact solution of the Klein-Gordon equation in terms of the local Heun function.
The local solutions at the event horizon asymptotically approach to the in/outgoing modes, whereas the local solutions at the infinity asymptotically approach to the growing/decaying modes.
The exact solution allows us to search the QN modes and clarify their rich structures.
We also developed a method to extract in/outgoing waves near the AdS boundary, which we summarize below, and calculated the greybody factor.
We then investigated the Hawking radiation of Kerr-AdS$_5$ black hole.

We calculated QN modes by requiring the ingoing boundary condition at the event horizon and the decaying boundary condition at the conformal infinity.
We checked the flow of QN modes with respect to the spin parameters.
Depending on the value of total magnetic quantum number, the degeneracy of the QN modes is broken and multiple branches show up for nonvanishing spin parameters, in parallel to the Zeeman splitting.
Furthermore, in addition to the known sequences of QN modes, we found the existence of the purely imaginary modes. 
A caveat is that the purely imaginary modes found in the present paper may or (partially) may not be the QN modes since there is some subtlety on the purely imaginary modes \cite{MaassenvandenBrink:2000iwh,Cook:2016fge,Cook:2016ngj}, which requires an independent study.

The purely imaginary modes are aligned on the negative imaginary axis at almost equal intervals. 
In the near extremal case, the interval is approximately given by $2\pi T_+$, where $T_+$ is the Hawking temperature at the event horizon.
The purely imaginary modes have also been found in different context, where the critical behavior has been observed, i.e., the purely imaginary mode goes to $-i\infty$ at some threshold parameter.
In our calculation, such a critical behavior was not observed for the equally rotating Kerr-AdS$_5$ with the common spin parameter down to $a=0.13$.
Nevertheless, we stress that it does not necessarily imply that the critical behavior does not exist for more general setup such as a Kerr-Newman-AdS$_5$ black hole.  
A deeper understanding of the behavior and physical meaning of the purely imaginary modes for the asymptotically AdS spacetime requires a further investigation.
It would be also intriguing to clarify their role in the context of gauge/gravity and/or fluid/gravity correspondence.

We explored the Hawking radiation of the Kerr-AdS$_5$ black hole by calculating the greybody factor for the test massive scalar field.
Here, as a first step toward the rigorous evaluation of the evaporation rate by taking into account other fields such as photon, graviton, and neutrinos, we focused on the contribution from the Hawking radiation of the scalar field, and demonstrate the computation of the evaporation rates for the mass and spins of the black hole.
Our result indicates that the larger angular momentum evaporates faster. 
It seems that the values of the two angular momenta may converge to the equal rotation case at least in the regime which we focused on. 
Of course, to grasp the comprehensive property of the contribution of the scalar field to the evaporation of Kerr-AdS$_5$ black hole 
such as how the initial state of the black hole determines spin-up/down by Hawking evaporation, 
it is necessary to analyse those evaporation speeds in a wide range over $a_1$-$a_2$ parameter space. 
We leave this analysis for future work.

Let us summarize our method for the computation of the greybody factor and the evaporation rates established in the present paper.
First, one needs to check the ``in/outgoingness'', i.e., if the exact solutions in terms of the Heun function describe the in/outgoing waves around the local minimum of the effective potential by the following prescription:
\begin{enumerate}
\item Find $u_{\rm b}$, which is the minimum positive root of $V'_{\rm eff}(u)=0$, where $u=\omega L^2/r$. 
The radius $r_{\rm b}=\omega L^2/u_{\rm b}$ is the location of the local minimum of the effective potential at the vicinity of the AdS boundary. 
\item Check if $V_\text{eff}(u_\text{b}) < 0$ and if the exact solution in terms of the local Heun function describes in/outgoing waves at the vicinity of $u=u_{\rm b}$.
As rough criteria, one can check if $\epsilon \ll 1$ and $\eta \ll 1$ are satisfied as given in \eqref{criteria}.
\item More precise criterion is to check if the imaginary parts of the phase~\eqref{eq:Imlog} of the exact solutions $R_\pm$ are linear in $u$.
\end{enumerate}
Our method with the Heun function applies to a wide parameter region, complementary to the previous work for the massless scalar field. 
Ultimately, it is safe to check if the local Heun function describes in/outgoing waves for all $(\ell,m_1,m_2)$ and $\omega$ of interest.
However, understanding a typical tendency of the parameters for the criteria \eqref{criteria}, one could skip to check most of the parameter region and only check the most dangerous parameter set. We note
that the parameter $\eta$ is prone to increase as $\ell$ increases or $\omega$ decreases (see Figs.~\ref{fig:region} and \ref{fig:eta_ell}). 
Also, it seems that $\eta$ depends on $m_1,m_2$ weakly, so it would be sufficient to check the in/outgoingness for some arbitrary $m_1,m_2$.
One can then proceed to the calculation of the Hawking evaporation:
\begin{enumerate}
\item Set some fiducial truncation of $\ell$, say $\ell_{\rm max}$.  
Find the smallest frequency $\omega_{\rm min}$ for $\ell=\ell_{\rm max}$ for which the exact solutions describe in/outgoing waves.
\item Calculate the power spectrum of the evaporation rates with the greybody factor, and check if the power spectrum is sufficiently small at $\ell_{\rm max}$ and $\omega_{\rm min}$.
\item If the contribution at $\ell_{\rm max}$ and $\omega_{\rm min}$ is nonnegligible, go back to the first step, take larger $\ell_{\rm max}$, and iterate the steps until the exact solutions remain to describe in/outgoing waves at sufficiently large $\ell_{\rm max}$ and small $\omega_{\rm min}$, where the contribution of the power spectrum to the evaporation rate is negligible.
\item After the above iteration converges, one can integrate the power spectrum over $\omega$ and obtain the evaporation rates.  
\end{enumerate}
Although the in/outgoingness is not quantitatively defined, one can quantify it by using $\epsilon, \eta$ or the imaginary part of the phase of the exact solutions. 
Of course, if one defines some threshold for in/outgoingness, $\ell_{\rm max}$ and $\omega_{\rm min}$ would change depending on the threshold.
Nevertheless, such a change is not crucial so long as the contribution at $\ell_{\rm max}$ and $\omega_{\rm min}$ to the evaporation rates is sufficiently small.  
The in/outgoingness is more manifest for smaller $\ell$, from which the dominant contribution to the evaporation rates comes.
Therefore, the above strategy is robust.
While we focused on the contribution from the scalar field, if the master equation of other fields such as photon, graviton, and neutrinos also take the form of the Heun equation, it is straightforward to apply our method to those fields.

In addition to the ones mentioned above, there are several other directions to extend the present work.
While we calculated the QN modes of the massive scalar field in Kerr-AdS$_5$ spacetime by imposing the ingoing and the Dirichlet boundary condition at the event horizon and the conformal infinity, respectively, one can also impose other boundary condition known as Robin boundary condition at the conformal infinity in more general cases~\cite{Ishibashi:2004wx,Dappiaggi:2017pbe,Katagiri:2020mvm, Katagiri:2022qje}. 
It would be interesting to apply our formalism to the QN modes with other boundary condition. 
Our method presented in this paper may be useful for relatively detailed computation of gauge/gravity correspondence since we 
employ the exact solution of the perturbation equation and one can obtain QN modes, greybody factor with high precision for a wide range of parameters. 
Moreover, it is also interesting to generalize our formalism to Kerr-Newman-AdS$_5$ spacetime, and to explore QN modes and Hawking evaporation.
Additionally, the analytic expression of the coefficients for the asymptotic form of the scalar field may be useful for an application in the context of gauge/gravity correspondence such as the GKP-Witten relation.
We leave these investigations for future work.

\acknowledgments

The authors thank Masataka Matsumoto, Shingo Kukita, and Mikio Nakahara for valuable comments on the gauge/gravity correspondence and the wave scattering problem for the greybody factor. 
The authors also thank Naritaka 
Oshita for discussion about 
the quasinormal modes of the Kerr-AdS$_5$ spacetime.
H.M.\ was supported by Japan Society for the Promotion of Science (JSPS) Grants-in-Aid for Scientific Research (KAKENHI) Grant Numbers JP18K13565 and JP22K03639.

\bibliographystyle{JHEPmod}
\bibliography{main}

\end{document}